\documentclass[fleqn,10pt]{wlscirep}
 \pdfoutput=1
\usepackage{setspace}
\usepackage[utf8]{inputenc}
\usepackage{tabu}
\usepackage{lineno}
\usepackage{arydshln}
\usepackage{amsmath,amssymb}
\usepackage{multicol,multirow}
\usepackage{etoolbox}
\usepackage{breqn}

\BeforeBeginEnvironment{dmath}{\begin{nolinenumbers}}%
\AfterEndEnvironment{dmath}{\end{nolinenumbers}}
\usepackage{booktabs}
\usepackage{graphicx}
\usepackage{subfigure}
\usepackage{epstopdf}
\usepackage{mathtools, cuted}
\usepackage{bm}
\usepackage[]{algorithm2e}

\newenvironment{remark}[1][Remark.]{\begin{trivlist}
\item[\hskip \labelsep {\bfseries #1}]}{\end{trivlist}}

\title{Beyond Trans-dimensional RJMCMC: Application to Impulsive Data Modeling}

\author[1]{Oktay Karakuş}
\author[2]{Ercan E. Kuruoğlu}
\author[1]{Mustafa A. Altınkaya}
\affil[1]{İzmir Institute of Technology (IZTECH), Electrical-Electronics Engineering, İzmir, Turkey}
\affil[2]{ISTI-CNR, via G. Moruzzi 1, 56124, Pisa, Italy}

\keywords{Keyword1, Keyword2, Keyword3}

\begin{abstract}
\textit{Reversible jump Markov chain Monte Carlo} (RJMCMC) is a Bayesian model estimation method which has been used for trans-dimensional sampling. In this study, we propose utilization of RJMCMC beyond trans-dimensional sampling. This new interpretation, which we call \textit{trans-space RJMCMC}, reveals the undiscovered potential of RJMCMC by exploiting the original formulation to explore spaces of different classes or structures. This provides flexibility in using different types of candidate classes in the combined model space such as spaces of linear and nonlinear models or of various distribution families. As an application for the proposed method, we have performed a special case of trans-space sampling, namely \textit{trans-distributional} RJMCMC in impulsive data modeling. In many areas such as seismology, radar, image, using Gaussian models is a common practice due to analytical ease. However, many noise processes do not follow a Gaussian character and generally exhibit events too impulsive to be successfully described by the Gaussian model. We test the proposed method to choose between various impulsive distribution families to model both synthetically generated noise processes and real life measurements on \textit{power line communications} (PLC) impulsive noises and \textit{2-D discrete wavelet transform} (2-D DWT) coefficients.
\end{abstract}
\AfterPreamble{\hypersetup{
  pdfauthor={Oktay Karakus},
  pdftitle={Beyond Trans-dimensional RJMCMC},
  pdfsubject={publication},
  urlcolor=blue,
}}
\usepackage{ifpdf}
\begin{document}

\flushbottom
\maketitle
\thispagestyle{empty}
\section{Introduction}\label{sec:intro}
In various real-life modeling problems, we have limited prior information regarding which model family is more suitable for the problem. In such cases, a method that would allow one to choose between different model families on the fly would be useful, eliminating the need for modeling with each candidate model class separately and comparing. This provides computational gains especially when the number of parameters and candidate model classes are high. An example is the choice between different \textit{probability density function} (pdf) models for noise or signals.

The pdf estimation problem is a frequently encountered problem in signal processing and statistics, and their application fields such as in image processing and telecommunications. In communication systems, channel modelling has been an important issue so as to characterize the whole system. However, for most of the cases, performing a deterministic channel modelling might be impossible and to represent real life systems, statistical channel models are very important. In addition, in applications of noise reduction operations in image processing, power-line communication systems, etc. dealing with a suitable statistical model beforehand is also important for the methods to be developed. Despite this importance, estimating the correct (or suitable) probability distribution along with its parameters within a number of generic distribution models may necessitate testing each candidate in order to choose the best possible model for the observed data/noise.

General tendency is to model noise/data with a Gaussian process especially in communications, network modeling, digital images, due to its analytical ease. In the case of non-Gaussian impulsive noise/data, various model families exist, for example, Middleton Class A, Bernoulli-Gaussian, $\alpha$-Stable, Generalized Gaussian (GG), Student's t, etc. It has been reported in the literature that noise exhibits non-Gaussian and impulsive characteristics in application areas such as wireless communications \cite{bhatti2009impulsive,blackard1993measurements}, \textit{power line communications} (PLC) \cite{lin2013impulsive,alsusa2013dynamic}, \textit{digital subscriber lines} (xDSL) \cite{al2011impulsive,fantacci2010impulse}, image processing \cite{simoncelli1997statistical, achim2003sar} and seismology \cite{yue2015validation}.

\textit{Reversible jump Markov chain Monte Carlo} (RJMCMC) is a Bayesian model determination method which has had success in vast areas of applications since its introduction by Peter Green \cite{green1995}. Unlike the widespread MCMC algorithm, \textit{Metropolis-Hastings} (MH), RJMCMC allows one to search in solution spaces of different dimensions which has been the main motivation for its use up to date. Classical applications of RJMCMC are model selection in regression and mixture processes \cite{troughton1997reversible, ehlers2004bayesian, eugri2010bayesian, richardson1997bayesian, viallefont2002bayesian, salas2009finite}. Unlike the classical applications in the literature, the original formulation of RJMCMC in \cite{green1995} permits a wider interpretation than just exploring the models with different dimensions. As an example of the applicability of RJMCMC beyond model dimension selection: it was utilized to learn polynomial autoregressive (PAR) \cite{karakus2015PAR}, polynomial moving average (PMA) \cite{karakus2016PMA} and polynomial autoregressive moving average (PARMA) \cite{karakus2017PARMA} processes and identification of Volterra system models \cite{karakucs2017bayesian} by exploring linear and nonlinear model spaces in preliminary work by the authors.

This paper contributes to the literature with a new interpretation on RJMCMC beyond trans-dimensional sampling, which we call \textit{trans-space RJMCMC}. The proposed method uses RJMCMC in an unorthodox way and reveals its potential to be a general estimation method by performing the reversible jump mechanism between spaces of different model classes. To demonstrate this potential, we focus our attention on a more special but generic problem of choosing between different probability distribution families. The problem is a frequently encountered problem in signal processing and statistics, and their application fields such as in image processing and telecommunications.

In this paper, we propose a Bayesian statistical modeling study of impulsive noise/data by estimating the probability distribution among three conventional impulsive distributions families: symmetric $\alpha$-Stable (S$\alpha$S), GG and Student's $t$. Other than identifying the distribution family, the proposed method estimates shape and scale parameters of the distribution. These distributions are the most popular statistical models in applications covering diverse areas such as wireless channel modeling, financial time series analysis, seismology, radar imaging.

We study the algorithm extensively on synthetic data providing statistical significance tests. In addition, as case studies, we look into two statistical modeling problems of actual interest impulsive noise on PLC channels and \textit{2-D discrete wavelet transform} (2-D DWT) coefficients. Particularly, PLC impulsive noise measurements in \cite{cortes2010analysis,lopes2013dealing} have been utilized in the simulations. Apart from this, statistical modeling for 2-D DWT coefficients have been performed on different kinds of images such as Lena, \textit{synthetic aperture radar} (SAR) \cite{SAR3}, \textit{magnetic resonance imaging} (MRI) \cite{MRI} and mammogram \cite{mammogram1}.

Rest of the paper is organized as follows: general definitions for trans-dimensional RJMCMC and the proposed method are discussed in Section \ref{sec:RJMCMC}. Section \ref{sec:RJMCMCinImp} reviews three distribution families and describes the impulsive data modeling scheme of the proposed method. Experimental studies for synthetically generated noise processes and for real applications are explained in Section \ref{sec:sim}. Section \ref{sec:conclusion} draws conclusions on the results.

\section{Reversible jump MCMC}\label{sec:RJMCMC}
RJMCMC has been first introduced by Peter Green in \cite{green1995} as an extension of MCMC to a model selection method. Green, Green firstly derives the condition for the satisfaction of detailed balance requirements in terms of the Borel sets which the candidate models belong to. In the continuation of the derivation, he specializes his discussion to moves between spaces which differ only in dimensions and the general discussion is abandoned. In the follow up, to the best of our knowledge almost all publications utilized RJMCMC for model dimension selection. Popular use of RJMCMC is in linear parametric models such as \textit{autoregressive} (AR) \cite{troughton1997reversible}, \textit{autoregressive integrated moving average} (ARIMA) \cite{ehlers2004bayesian} and \textit{fractional} ARIMA (ARFIMA) \cite{eugri2010bayesian} and mixture models such as Gaussian mixtures \cite{richardson1997bayesian}, Poisson mixtures \cite{viallefont2002bayesian} and $\alpha$-stable mixtures \cite{salas2009finite}.

Apart from the popular applications above, RJMCMC has been used in other various applications such as detection of clusters in disease maps \cite{knorr2000bayesian}, graphical models based variable selection and automatic curve fitting \cite{lunn2009generic}, log-linear model selection \cite{dellaportas1999markov}, non-parametric drift estimation \cite{van2014reversible}, delimiting species using multilocus sequence data \cite{rannala2013improved}, random effect models \cite{oedekoven2016using}, generation of lane-accurate road network maps from vehicle trajectory data \cite{roeth2017extracting}.

In this study, our motivation is to propose a new interpretation on the classical RJMCMC beyond trans-dimensionality. The classical trans-dimensional RJMCMC of \cite{green1995} and the proposed method, \textit{trans-space} RJMCMC are discussed in the sequel.

\subsection{Trans-dimensional RJMCMC}
The standard MH algorithm \cite{Hastings70} accepts a transition from Markov chain state $x\in\mathcal{X}$ to $y\in\mathcal{X}$ with a probability of:

\begin{align}\label{equ:mh}
  A(x\rightarrow y) = \min \left\{ 1, \dfrac{\pi(y)q(x,y)}{\pi(x)q(y,x)} \right\}
\end{align}
where $\pi(\cdot)$ represents the target distribution and $q(y,x)$ refers to the proposal distribution from state $x$ to $y$.

RJMCMC, in the sense of trans-dimensional MCMC, generalizes MH algorithm by defining multiple parameter subspaces $\zeta_k$ of different dimensionality \cite{green1995}. This is only achieved by defining different types of moves between subspaces providing that the detailed balance is attained. For this condition to hold, a reverse move from state $y$ to $x$ should be defined and dimension matching should be satisfied between parameter subspaces.

Assume that we propose a move $m$ with probability $p_m$ from a Markov chain state $\kappa$ to $\kappa'$ each of which has parameter vectors $\theta\in\zeta_1$ and $\theta'\in\zeta_2$, respectively with different dimensions. The move $m$ is reversible and its reverse move $m^{\text{R}}$ is proposed with a probability $p_{m^{\text{R}}}$.  The general detailed balance condition can be stated as:

\begin{align}\label{equ:detbal}
  \pi(\kappa) q(\kappa', \kappa) A(\kappa \rightarrow \kappa') = \pi(\kappa') q(\kappa, \kappa') A(\kappa' \rightarrow \kappa),
\end{align}
where proposal distribution $q(\cdot)$ is directional and includes the probabilities of both the move itself and the proposed parameters. Then, the general expression for the acceptance ratio in (\ref{equ:mh}) turns into \cite{green1995}:

\begin{dmath}\label{equ:rjmcmc_alpha}
A(\kappa \rightarrow \kappa') = \min \left\{ 1, \dfrac{\pi(\kappa') p_{m^{\text{R}}} \chi_2(\mathbf{u'})}{\pi(\kappa) p_m \chi_1(\mathbf{u})} \left| \dfrac{\partial (\theta', \mathbf{u'})}{\partial (\theta, \mathbf{u})} \right| \right\},
\end{dmath}
where $\chi_1(\cdot)$ and $\chi_2(\cdot)$ are the distributions for the auxiliary variable vectors $\mathbf{u}$ and $\mathbf{u'}$, respectively which are required to provide dimension matching for the moves $m$ and $m^{\text{R}}$. The term $\left| \frac{\partial (\theta' \mathbf{u'})}{\partial (\theta, \mathbf{u})} \right|$ is the magnitude of the Jacobian.

In each RJMCMC run, the standard Metropolis-Hastings algorithm is applied in moves within the same dimensional models, which is called as \textit{life} move. Sampling is performed in a single parameter space and there is no dimension change in life move. For trans-dimensional transitions between models, moves such as \textit{birth}, \textit{death}, \textit{split} and \textit{merge} are performed which require the creation or the deletion of new variables corresponding to the increased or decreased dimension. Green handles the dimension changing moves as variable transformations and defines a dummy variable to match dimensions which provides a square Jacobian matrix that can be used to update the acceptance ratio easily.

\subsection{Trans-space RJMCMC}
In spite of RJMCMC's use in trans-dimensional cases, the original formulation in \cite{green1995} holds a wider interpretation than just sampling between spaces of different dimensions. In the beyond trans-dimensional RJMCMC point of view, the main requirements of RJMCMC stated by Green are still valid with one exception, that is, a change in parameter space definition.

In the original formulation, Green firstly derives the condition for the satisfaction of detailed balance requirements in terms of the Borel sets which the candidate models belong to. In the continuation of the derivation, he specializes his discussion to moves between spaces which differ only in dimensions and the general discussion is abandoned. However, the parameter vectors in (\ref{equ:detbal}) may belong to Borel sets which differ not only in their dimensions but also in the generic models they belong to. Thus, the algorithm can be used for much more generic implementations.

Notwithstanding, this general interpretation should be taken with caution to have a useful method. Particularly, the Borel sets should be \emph{related} somehow, which can be conveniently set by \emph{matching a common property (e.g. norm)} in defining the spaces. Defining proposals in this way will provide sampling more efficient candidates and help algorithm to converge faster. As an example, model transitions can be designed to provide fixed first ordered moments between spaces. Thus, this moment based approach provides a more efficient way to explore all the candidate models within the combined space. Carrying the trained information to a new generic model space is very crucial in this framework. Otherwise, the algorithm would start to train from scratch repeatedly each time it changes states and sampling across unrelated spaces would not give us a computational advantage. In that case, one could solve for different spaces separately and compare the final results to choose the best model. Two examples one can think of firstly, are:
\begin{enumerate}
\item $\kappa$ might correspond to a linear parametric model such as AR while $\kappa'$ might correspond to a nonlinear model such as Volterra AR.
\item $\kappa$ might correspond to a pdf $p_A$ with certain distribution parameters while $\kappa'$ might correspond to another pdf $p_B$ with some other distribution parameters.
\end{enumerate}

To this end, we define a combined parameter space $\varphi=\bigcup_k \varphi_k$ for $k > 1$. Assume that a move $M$ from Markov chain state $x\in\varphi_1$ to $x'\in\varphi_2$ is defined and Borel sets $A\subset\varphi_1$ and $B\subset\varphi_2$ are related with a set of functions each of which are invertible. Particularly, for any Borel sets in both of the spaces, $\varphi_1$ and $\varphi_2$, functions $h_{12}:A\mapsto B$ and $h_{21}:B\mapsto A$ can be defined by matching a common property of the spaces. For generality, if the proposed move requires matching the dimensions, auxiliary variables $\mathbf{u}_1$ and/or $\mathbf{u}_2$ can be drawn from proper densities $Q_1(\cdot)$ and $Q_2(\cdot)$, respectively. Otherwise, one can set $\mathbf{u}_1$ and $\mathbf{u}_2$ to $\emptyset$. Please note that the dimensions of the parameter spaces at both sides of the transitions can be different or the same and reversible jump mechanism is still applicable.

Consequently, although the candidate spaces are of different classes, since the Borel sets are defined as to be related, the assumption of Green still holds for a symmetric measure $\xi_m$ and densities for joint proposal distributions, $\pi(\cdot)q(\cdot, \cdot)$, can be defined with respect to this symmetric measure by satisfying the equilibrium in (\ref{equ:detbal}). Thus, the acceptance ratio can be written as:

\begin{dmath}\label{equ:rjmcmc_alpha2}
A(x \rightarrow x') = \min \left\{ 1, \dfrac{\pi(x') p_{M^{\text{R}}} Q_2(\mathbf{u}_2)}{\pi(x) p_M Q_1(\mathbf{u}_1)} \left| \dfrac{\partial h_{12}(\theta_1, \mathbf{u}_1)}{\partial (\theta_1, \mathbf{u}_1)} \right| \right\}.
\end{dmath}
where $M^R$ is the reverse move of $M$ and $p_M$ and $p_{M^{\text{R}}}$ represent the probabilities of the moves. The Jacobian term appears in the equation as a result of the change of variables operation between spaces.

Here we recall that in our previous works \cite{karakus2015PAR,karakus2016PMA, karakus2017PARMA, karakucs2017bayesian}, we have performed model estimation studies with RJMCMC for Volterra based nonlinear models PAR, PMA and PARMA as well as an identification study of Volterra system models. In these studies, RJMCMC has been utilized to explore the model spaces of linear and nonlinear models in polynomial sense instead of performing a model order selection study in a single linear model space. Hence, we add a few concluding remarks.

\begin{remark}[Remark1.]
We are going to name this new utilization on RJMCMC as \emph{trans-space} rather than \textit{trans-dimensional}. Trans-space RJMCMC reveals a general framework for exploring the spaces of different generic models whether or not their parameter spaces are of different dimensionality. Consequently, trans-dimensional case is a subset of trans-space transitions.
\end{remark}

\begin{remark}[Remark2.]
Trans-space RJMCMC requires to define new types of moves due to the need for more detailed operations than, e.g. just being birth, death, split and merge of the parameters. These moves will be named as \textit{between-space moves} and may include both \textit{birth} and \textit{death} of the parameters at the same time or a norm based mapping between the parameter spaces. \textit{Switch} move (firstly proposed for Volterra system identification study \cite{karakucs2017bayesian}) will be proposed as a between-space move, which performs a switching between the candidate spaces of the generic model classes.
\end{remark}

\begin{remark}[Remark3.]
As a special case of trans-space sampling, the proposed method can be used to explore the spaces of different distribution families. Therefore, this special case will be named as \textit{trans-distributional}.
\end{remark}

\section{Trans-distributional RJMCMC for Impulsive Distributions}\label{sec:RJMCMCinImp}
In this study, we have applied RJMCMC to problems in which a stochastic process, $\mathbf{x}$, is given whose impulsive distribution is to be found. For this purpose, we define a reversible jump mechanism which estimates the distribution family among three impulsive distribution families, namely, S$\alpha$S, GG and Student's $t$.

These three families cover many different noise modeling studies as stated in the above sections. All of them include Gaussian distribution as a special member, and many real life noise measurements can be modelled with these distribution families. For example, S$\alpha$S family has various demonstrated application areas such as PLC \cite{laguna2015use}, SAR imaging \cite{achim2003sar}, near optimal receiver design \cite{kuruoglu1998near}, modelling of counterlet transform subbands \cite{sadreazami2014study}, seismic amplitude data modelling \cite{yue2015validation}, as noise model for molecular communication \cite{farsad2015stable}, reconstruction of non-negative signals \cite{tzagkarakis2010greedy} (Please see \cite{nolan2010bibliography} and references therein for detailed applications).

GG distributions have found applications in wavelet based texture retrieval \cite{do2002wavelet}, image modelling in terms of Markov random fields \cite{bouman1993generalized}, multicomponent texture discrimination in color images \cite{verdoolaege2011geodesics}, wheezing sound detection \cite{le2009wheezing}, modelling sea-clutter data \cite{novey2010complex}.

Student's $t$ distribution is an alternative to Gaussian distribution especially for small populations where the validity of central limit theorem is questionable. Student's $t$ distribution has been used in applications of finance \cite{patton2006modelling, engle1986modelling}, full-waveform inversion of seismic data \cite{aravkin2011robust}, independent vector analysis for speech separation \cite{liang2013independent}, medical image segmentation \cite{nguyen2012robust}, growth curve modelling \cite{zhang2013bayesian}.

One might argue that training separate MCMC samplers for each of the seemingly irrelevant distribution families and comparing their modelling performances afterwards would be computationally more advantageous. However, in cases when the number of candidate models is not known or dramatically large, implementing a single Markov chain via RJMCMC could be simpler. In addition, when the number of models are small, one can not conclude that parallel MCMC approach would be a better choice than RJMCMC and this requires an analysis. By efficiently choosing the proposal distributions, the advantage of incorporating reversible jump mechanism can be extended to searching several distribution families which will be described in the sequel.

In the literature, RJMCMC usage in this problem has been limited and it has been used to be examples of trans-dimensional approach deciding between two specific distributions \cite{hastie2012model, barker2013bayesian}. Particularly, when modelling count data, reversible jump mechanism has been applied to choose between Poisson and negative binomial distributions in \cite{hastie2012model}. This study deals with the question whether the count data is over-dispersed relative to Poisson distribution. In \cite{barker2013bayesian} an approach which is a combination of Gibbs sampler and RJMCMC has been used to decide between Poisson and geometric distributions by using a universal parameter space called ``palette". Both of the studies have utilized RJMCMC in distribution estimation; however, the number of candidate distributions was limited to two. Moreover, in both of the studies, Poisson distribution is a special member of the distribution families in question (or, there is a direct relation between Poisson and negative binomial or geometric distributions), hence, the methods in these studies can be handled with a single family search.

The proposed method, \emph{trans-distributional} RJMCMC, is much more general than the examples above and aims to fit a distribution to a given process $\mathbf{x}$ among various distributions by identifying the distribution's family and estimating its shape and scale parameters. Two types of between-class moves have been defined, namely \emph{intra-class-switch} and \emph{inter-class-switch}. These moves propose model class changes \textit{within} and \textit{between} probability distribution families, respectively.
\subsection{Impulsive Distribution Families} \label{sec:distfamilies}
\subsubsection{Symmetric $\alpha$-Stable Distribution Family}
There is no closed form expression for probability density function (pdf) of S$\alpha$S distributions except for the special cases of Cauchy and Gaussian. However, its characteristic function, $\varphi(x)$, can be expressed explicitly as:

\begin{align}\label{equ:aS_CF}
\varphi(x) = \exp(j\delta x - \gamma|x|^{\alpha})
\end{align}
where $0<\alpha \leq 2$ is the characteristic exponent, \emph{a.k.a. shape parameter}, which controls the impulsiveness of the distribution. Special cases Cauchy and Gaussian distributions occur when $\alpha=1$ and $\alpha=2$, respectively. $-\infty<\delta<\infty$ represents the \emph{location parameter}. The $\gamma>0$ provides a measure of the dispersion which is the \emph{scale parameter} expressing the spread of the distribution around $\delta$.

\subsubsection{Generalized Gaussian Distribution Family}
The univariate GG pdf can be defined as:

\begin{align}\label{equ:GG_pdf}
f(x) = \dfrac{\alpha}{2\gamma\Gamma(1/\alpha)} \exp\left(-\left(\dfrac{|x-\delta|}{\gamma}\right)^{\alpha}\right)
\end{align}
where $\Gamma(\cdot)$ refers to the gamma function, $\alpha>0$ is the shape parameter, $-\infty<\delta<\infty$ represents the location parameter and the $\gamma>0$ is the scale parameter. GG family has well-known members such as Laplace, Gauss and uniform distributions for $\alpha$ values of 1, 2 and $\infty$, respectively.

\subsubsection{Student's $t$ Distribution Family}
The univariate symmetric Student's $t$ distribution family is an impulsive distribution family with parameters, $\alpha>0$ which is the number of degrees of freedom, \emph{a.k.a shape parameter}, the location parameter $-\infty<\delta<\infty$ and the scale parameter $\gamma>0$. Its pdf can be defined as:

\begin{align}\label{equ:t_pdf}
f(x) = \dfrac{\Gamma\left(\dfrac{\alpha+1}{2}\right)}{\Gamma(\alpha/2) \gamma \sqrt{\pi \alpha}} \left( 1+\dfrac{1}{\alpha} \left( \dfrac{x - \delta}{\gamma} \right)^2 \right)^{-((\alpha+1)/2)}.
\end{align}

Special members of the symmetric Student's $t$ distribution family are Cauchy and Gauss which are obtained for shape parameter values of $\alpha=1$ and $\alpha=\infty$, respectively.
\subsection{Parameter Space}\label{sec:paramspace}
RJMCMC construction for impulsive data modeling begins by firstly defining the parameter space. Parameter space has been defined on the common parameters for all three distribution families. These are: \emph{shape, scale} and \emph{location} parameters ($\alpha, \gamma$ and $\delta$, respectively). In addition to them, \emph{the family identifier}, $k$, which defines the estimated distribution family has been added to the parameter space. The $k$ values of the distributions S$\alpha$S, $\text{GG}$ and Student's $t$ are 1, 2 and 3, respectively. Therefore, the parameter vector $\theta$ can be formed as: $\theta = \{ k, \alpha, \delta, \gamma \}$.

In this study, the observed data from all three families are assumed to be symmetric around the origin for simplicity. Therefore, $\delta$, is set to 0 and its effect will be invisible in the simulations. Consequently, parameter vector $\theta$ is reduced to: $\theta = \{ k, \alpha, \gamma \}$.

\subsection{Hierarchical Bayesian Model}\label{sec:bayeshier}
The target distribution, $f(\theta|\mathbf{x})$, can be decomposed to likelihood times priors due to Bayes Theorem as:

\begin{align}\label{postlhdpri}
f(\theta|\mathbf{x}) \propto f(\mathbf{x}|k, \alpha, \gamma) f(\alpha|k) f(k) f(\gamma).
\end{align}
where $f(\mathbf{x}|k, \alpha, \gamma)$ represents the likelihood and $f(\alpha|k), f(k)$, and $f(\gamma)$ are the priors.

\subsection{Likelihood}\label{sec:lhd}
We assume that the stochastic process $\mathbf{x}$ with a length of $n$ comes from one of the distributions in candidate families (S$\alpha$S, GG and Student's $t$). Then, the likelihood corresponds to a pdf from one of these distributions:

\begin{align}\label{equ:likelihood}
f(\mathbf{x}|k, \alpha, \gamma) &= \left\{
  \begin{array}{ll}
    \prod_{i=1}^{n} \text{S$\alpha$S}(\gamma), &k=1 \\
    \prod_{i=1}^{n} \text{GG}_{\alpha}(\gamma), &k=2 \\
    \prod_{i=1}^{n} t_{\alpha}(\gamma), &k=3
  \end{array} \right .
\end{align}

\subsection{Priors}\label{sec:priors}
Priors have been selected as the following:

\begin{align}\label{equ:priors}
  f(\gamma) &=  \mathcal{IG}(a, b),\\
  f(k) &=  \mathbb{I}_{\{1/3, 1/3, 1/3\}} \quad \text{for } k = 1, 2, 3,\\
  f(\alpha|k) &= \left\{
  \begin{array}{ll}
    \mathcal{U}(0, 2) & k=1, \\
    \mathcal{U}(0, \alpha_{\text{max,GG}}) & k=2, \\
    \mathcal{U}(0, \alpha_{\text{max},t}) & k=3,
  \end{array} \right.
\end{align}
where $a$ and $b$ represent the hyperparameters for scale parameter and they are generally selected as to take small values such as $1, 0.1$ in the literature. The upper bounds for the shape parameters of $\text{GG}$ and Student's $t$ distributions have been defined as $\alpha_{\text{max,GG}}$ and $\alpha_{\text{max},t}$, respectively.

Choosing an inverse gamma prior for scale parameter is a general practice especially for Gaussian problems. Due to the lack of information about conjugate priors for distributions other than the Gaussian case and since Gaussian distribution is common for all three families, an inverse gamma conjugate prior for scale parameters has been chosen for simplicity. Furthermore, all families are equiprobable \emph{a priori} and shape parameter is uniformly distributed between lower and upper bounds.

\subsection{Model Moves}\label{sec:moves}
Two RJMCMC model moves have been defined in order to perform trans-distributional transitions discussed in the previous sections. These are: \emph{life} and \emph{switch} moves. Life move performs classical MH algorithm to update $\gamma$. Switch move performs exploring the other distribution spaces. For this purpose, two types of switch moves have been defined: \textit{intra-class-switch} and \textit{inter-class-switch}. Intra-class-switch performs exploring the distributions in the same family, while inter-class-switch explores spaces of different families. At each RJMCMC iteration, one of the moves is chosen with probabilities $P_{\text{life}}, P_{\text{intra-cl-sw}}$ and $P_{\text{inter-cl-sw}}$, respectively.

\begin{figure}[ht!]
  \centering
  \includegraphics[width=.8\linewidth]{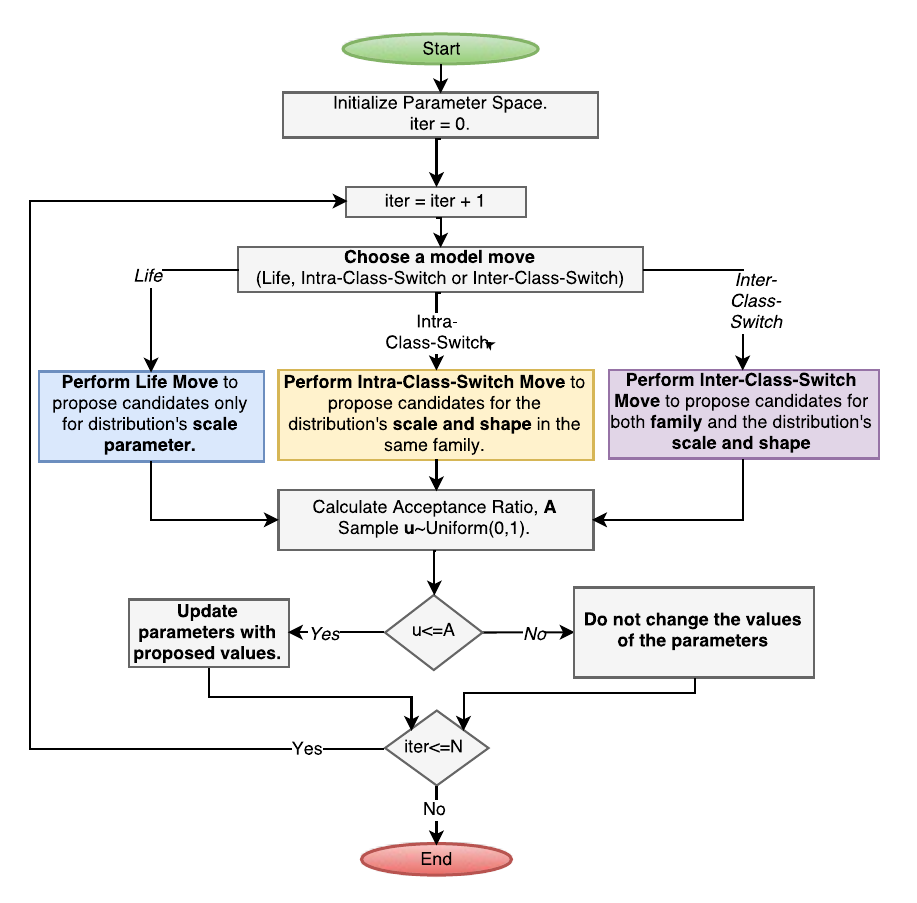}
  \caption{Flow Diagram for the Proposed method.}\label{fig:flow}
\end{figure}

In Figure \ref{fig:flow} the flow diagram of the proposed method is depicted where the parameter $N$ refers to the maximum number of iterations. The details about the steps of the selected moves are discussed in the sequel.

\subsubsection{Life Move}
\emph{Life} move defines a transition from parameter space $(k, \alpha, \gamma)$ to $(k', \alpha', \gamma')$ and only proposes a candidate for the scale parameter, $\gamma$ ($\alpha'=\alpha$ and $k'=k$). The proposal distribution for scale parameter $\gamma'$ has been chosen as:

\begin{align}\label{equ:lifeproposal}
  q(\gamma'|\gamma) = \mathcal{TN}(\gamma, \xi_{scale}) \quad \text{for interval } (0, \gamma+1]
\end{align}
where $\mathcal{TN}(\gamma, \xi_{scale})$ refers to a Gaussian distribution where its mean $\gamma$ is the last value of the scale parameter, and its variance is $\xi_{scale}$ and is truncated to lie within the interval of $(0, \gamma+1]$ afterwards by rejecting samples outside this interval. This truncation procedure aims to satisfy the condition $\gamma>0$ and forces candidate proposals not to lie far from the last value of $\gamma$. Hence, the resulting acceptance ratio for life move is:

\begin{align}\label{equ:lifeaccratio}
  A_{\text{life}} = \min \left\{ 1, \dfrac{f(\mathbf{x}|k', \alpha', \gamma')}{f(\mathbf{x}|k, \alpha, \gamma)} \dfrac{f(\gamma')}{f(\gamma)} \dfrac{q(\gamma|\gamma')}{q(\gamma'|\gamma)} \right\}
\end{align}

\subsubsection{FLOM Based Proposals for $\gamma$ Transitions}\label{sec:FLOMBased}
As mentioned earlier in this paper, using a common feature among the candidate model spaces for the transition to be made will provide efficient proposals and is important in order to link the subspaces of different classes. Assume we have two candidate families parameter vectors of which belong to Borel sets, $\mathcal{A}$ and $\mathcal{B}$, respectively. Providing fixed order norm for both of the Borel sets, the transition (e.g. $h:\mathcal{A}\mapsto\mathcal{B}$) from one set to another carries the information in the same direction which has been already learned at the most recent Borel set. Considering the convergence and mixing of the algorithm, such an approach is very important to determine the transition process between generic distribution models, whether within the family or between families.

When dealing with distribution estimation problems, moments with various orders, $p$ have been defined for all distribution families. Moments of $t$ and GG families have been defined at any orders for $p>0$ and there are no restrictions on values of $p$. However, moments of the S$\alpha$S family have been defined subject to the constraint of $p < \alpha$. This constraint makes it possible to use the absolute \textit{fractional lower order moments (FLOMs)} which has been also used in the parameter estimation methods of the S$\alpha$S family. By taking into consideration of the facts that absolute FLOM expressions are defined for all impulsive families, and their success in parameters estimation studies of the S$\alpha$S distributions, using an absolute FLOM based approach helps to construct a reversible jump sampler between different impulsive families, by linking the candidate distributions through absolute FLOM.

In impulsive data modelling study in this study, absolute FLOM-based approach will be used for the proposals of the $\gamma$ parameter. In particular, to perform sampling between related subspaces and generate efficient proposals on scale parameter $\gamma$, an absolute FLOM-based method has been used. The newly proposed scale parameter, $\gamma'$, is calculated via a reversible function, $g(\cdot)$ (or $w(\cdot)$), which provides equal absolute FLOMs with order $p$ for both the most recent and candidate distribution spaces. Thus, proposals on $\gamma$ carry the learned information to the candidate space via absolute FLOMs.

Absolute FLOMs are defined only for $p$ values lower than alpha for the case of S$\alpha$S distributions. Moreover, there are several studies which suggest near-optimum values for FLOM order $p$ in order to estimate the scale parameter of S$\alpha$S distributions. \cite{tsihrintzis1996fast} suggests $p=\alpha/4$ and \cite{ma1995parameter} suggests $p=0.2$. However, in \cite{kuruoglu2001density} it has been stated that decreasing $p$ for a fixed value of $\alpha$ (i.e. increasing $\alpha/p$), increases the estimation performance of $\gamma$ and \cite{kuruoglu2001density} suggests the choice $p=\alpha/10$. We use the value $p=\alpha/10$ in our simulations for all the distribution families.

For a given data, $\mathbf{x}$, in order to perform a transition from parameter space $\{ k, \alpha, \gamma\}$ to $\{ k', \alpha', \gamma'\}$ we assume that the absolute FLOM will be the same for both the most recent and candidate distribution spaces. In particular,

\begin{align}\label{equ:FLOM1}
  E_k(|\mathbf{x}|^p) = E_{k'}(|\mathbf{x}|^p)
\end{align}
where absolute FLOMs for all three candidate families can be defined as:

\begin{align}\label{equ:FLOM2}
    E_k(|\mathbf{x}|^p) &= \left\{
  \begin{array}{ll}
    C_{\alpha}(p, \alpha) \gamma^{p/\alpha} & k=1, \\
    C_{\text{GG}}(p, \alpha) \gamma^{p} & k=2, \\
    C_{t}(p, \alpha) \gamma^{p} & k=3,
  \end{array} \right.
\end{align}
where

\begin{align}
  C_{\alpha}(p, \alpha) &= \dfrac{\Gamma\left(\dfrac{p+1}{2}\right) \Gamma\left(\dfrac{-p}{\alpha}\right)}{\alpha \sqrt{\pi} \Gamma\left(\dfrac{-p}{2}\right)}2^{p+1}, \\
  C_{\text{GG}}(p, \alpha) &= \dfrac{\Gamma \left( \dfrac{p+1}{\alpha} \right)}{\Gamma(1/\alpha)}, \\
  \label{equ:FunctionCs}C_{t}(p, \alpha) &= \dfrac{\Gamma\left(\dfrac{p+1}{2}\right) \Gamma\left(\dfrac{\alpha-p}{2}\right)}{\sqrt{\pi} \Gamma\left(\dfrac{\alpha}{2}\right)}\alpha^{p/2}.
\end{align}

The candidate proposal, $\gamma'$, has been calculated via reversible functions which are derived by using the relations in (\ref{equ:FLOM1})-(\ref{equ:FunctionCs}) for each transition. These functions have been derived for both of the switch moves and are shown in Tables \ref{tab:LpIntra} and \ref{tab:LpInter}.

\subsubsection{Intra-Class-Switch Move}
RJMCMC performs a transition on shape and scale parameters in the same distribution family ($k'=k$) when an intra-class-switch move is proposed. The proposed shape parameter $\alpha'$ is sampled from a proposal distribution $q(\alpha'|\alpha)$. In addition, the candidate scale parameter $\gamma'$ is defined as a function $g(\alpha, \alpha', p, \gamma)$.

\begin{figure}[ht!]
\centering
\subfigure[]{\label{fig:proposaldist}
\includegraphics[width=.4\linewidth]{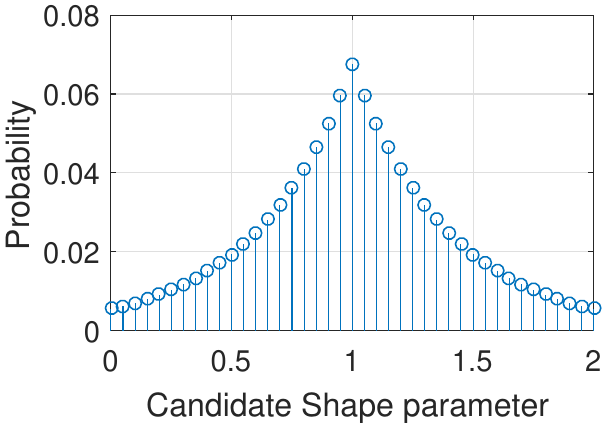}}
\centering
\subfigure[]{\label{fig:mappingfunc}
\includegraphics[width=.4\linewidth]{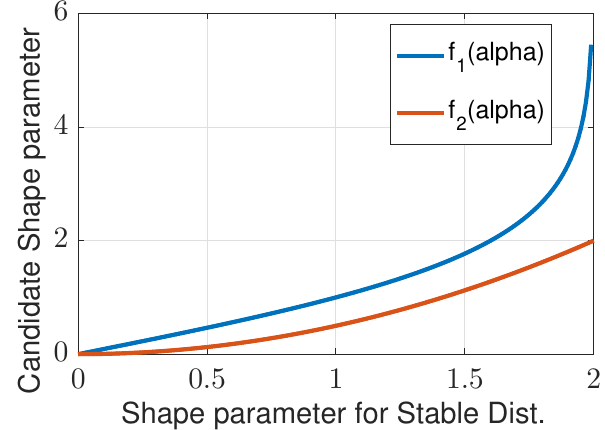}}
\caption{(a) - Proposal distribution, $q(\alpha'|\alpha)$ for intra-class-switch move $(\gamma=1, \Gamma=0.4)$. (b) - Mapping functions on shape parameter for inter-class-switch move}
\end{figure}

The $\gamma$ transition in this move has been defined as dependent on the newly proposed $\alpha'$ parameter. For this reason, a step is first performed on shape parameter $\alpha$ to propose $\alpha'$, and this is used to calculate the candidate scale parameter $\gamma'$. For the shape parameter $\alpha$ transition, a proposal distribution such as $q(\alpha'|\alpha)$ has been used. For this distribution, we first have assumed a symmetric distribution around the most recent $\alpha$ value. In addition, it has been preferred that the proposal distribution has heavier tails than Gaussian in order to make it possible to sample candidates much farther than the most recent $\alpha$ relative to the samples from the Gaussian distribution. Since the Laplace distribution is a distribution that satisfies all these conditions, the proposal distribution is chosen as a Laplace distribution. Due to the numerical calculation problems caused when $\alpha$ and $\alpha'$ are close to each other (i.e. $|\alpha - \alpha'| \leq 0.03$), we have decided to utilize a finite number of candidate distributions (i.e. a finite number of $\alpha$ values) and the space on $\alpha$ is discretized with increments of 0.05. That's why a discretized Laplace ($\mathcal{D}\mathcal{L}(\alpha, \Gamma)$) distribution where the location parameter of which is equal to the most recent shape parameter $\alpha$ and scale parameter is $\Gamma$, has been utilized. An example figure of the proposal distribution $q(\alpha'|\alpha)$ is shown in Figure \ref{fig:proposaldist}.

Importantly, our choice on the proposal distribution $q(\alpha'|\alpha)$ is not restrictive; any distribution other than Laplace can be selected as the proposal distribution (e.g. Gaussian like). However, different selections will cause the algorithms to perform differently.

Candidate scale parameter $\gamma'$ has been calculated via reversible functions, $g(\cdot)$, which are derived for intra-class-switch move by using the method in Section \ref{sec:FLOMBased}. Functions for each family are shown in Table \ref{tab:LpIntra}.

\begin{table}[ht!]
  \centering
  \caption{Intra-Class-Switch Details [$(k, \alpha, \gamma)\rightarrow(k', \alpha', \gamma')$]}\label{tab:LpIntra}
  \scriptsize
  \resizebox{.7\linewidth}{!}{\begin{tabular}{ccll}
     \hline
     Family & Degree, $p$ & $\gamma'=g(\alpha, \alpha', p, \gamma)$  & Jacobian, $|J|$\\
     \hline
     S$\alpha$S & $\alpha'/10$ & $\left(\dfrac{C_{\alpha}(p, \alpha)}{C_{\alpha}(p, \alpha')}\right)^{\alpha'/p} \gamma^{\alpha'/\alpha}$ & $\left(\dfrac{C_{\alpha}(p, \alpha)}{C_{\alpha}(p, \alpha')}\right)^{\alpha'/p} \dfrac{\alpha'}{\alpha} \gamma^{(\alpha' - \alpha)/\alpha}$ \\
     \hline%&&\\
     $\text{GG}$ & $\alpha'/10$ & $\left(\dfrac{C_{\text{GG}}(p, \alpha)}{C_{\text{GG}}(p, \alpha')}\right)^{1/p} \gamma $ & $\left(\dfrac{C_{\text{GG}}(p, \alpha)}{C_{\text{GG}}(p, \alpha')}\right)^{1/p}$\\
     \hline%&&\\
     $t$ & $\alpha'/10$ & $\left(\dfrac{C_{t}(p, \alpha)}{C_{t}(p, \alpha')}\right)^{1/p} \gamma$  & $\left(\dfrac{C_{t}(p, \alpha)}{C_{t}(p, \alpha')}\right)^{1/p}$\\
     %&&\\
     \hline
   \end{tabular}}
%\end{table}
%
%\begin{table}
  \centering
  \vspace{0.3cm}
  \caption{Inter-Class-Switch Details [$(k, \alpha, \gamma)\rightarrow(k', \alpha', \gamma')$]}\label{tab:LpInter}
  \scriptsize
  \resizebox{.7\linewidth}{!}{\begin{tabular}{ccll}
     \hline
     ($k \rightarrow k'$) & Degree, $p$ & $\alpha'=\psi(\alpha, k, k')$ & $\gamma'=w(\alpha, \alpha', p, \gamma)$\\
     \hline
     $1\rightarrow2$ & $\alpha'/10$ & $f_1(\alpha)=\dfrac{\alpha^2}{2}$ & $\left(\dfrac{C_{\alpha}(p, \alpha)}{C_{\text{GG}}(p, \alpha')}\right)^{1/p} \gamma^{1/\alpha}$\\
     \hline%&&&\\
          $1\rightarrow3$ & $\alpha'/10$ & $f_2(\alpha)=logit\left(\dfrac{\alpha+2}{4}\right)$ & $\left(\dfrac{C_{\alpha}(p, \alpha)}{C_{t}(p, \alpha')}\right)^{1/p} \gamma^{1/\alpha}$\\
     \hline%&&&\\
          $2\rightarrow1$ & $\alpha'/10$ & $f_1^{-1}(\alpha)$ & $\left(\dfrac{C_{\text{GG}}(p, \alpha)}{C_{\alpha}(p, \alpha')}\right)^{\alpha'/p} \gamma^{\alpha'}$\\
     \hline%&&&\\
          $2\rightarrow3$ & $\alpha'/10$ & $f_2(f_1^{-1}(\alpha))$ & $\left(\dfrac{C_{\text{GG}}(p, \alpha)}{C_{t}(p, \alpha')}\right)^{1/p} \gamma$\\
     \hline%&&&\\
          $3\rightarrow1$ & $\alpha'/10$ & $f_2^{-1}(\alpha)$  & $\left(\dfrac{C_{t}(p, \alpha)}{C_{\alpha}(p, \alpha')}\right)^{\alpha'/p} \gamma^{\alpha'}$\\
     \hline%&&&\\
          $3\rightarrow2$ & $\alpha'/10$ & $f_1(f_2^{-1}(\alpha))$ & $\left(\dfrac{C_{t}(p, \alpha)}{C_{\text{GG}}(p, \alpha')}\right)^{1/p} \gamma$\\
          %&&&\\
          \hline
   \end{tabular}}
\end{table}

Consequently, proposals for intra-class-switch move are;

\begin{align}\label{equ:inswtransitions}
  q(\alpha'|\alpha) &= \mathcal{D}\mathcal{L}(\alpha, \Gamma), \\
  \gamma' &= g(\alpha, \alpha', p, \gamma).
\end{align}

As a result of the details explained above, acceptance ratio for RJMCMC intra-class-switch move can be expressed as;

\begin{align}\label{equ:inswaccratio}
  A_{\text{intra-cl-sw}} = \min \left\{ 1, \dfrac{f(\mathbf{x}|k', \alpha', \gamma')}{f(\mathbf{x}|k, \alpha, \gamma)} \dfrac{f(\gamma')}{f(\gamma)} |J| \right\},
\end{align}
where $|J|$ is the magnitude of the Jacobian (See Table \ref{tab:LpIntra}).

\subsubsection{Inter-Class-Switch Move}
Different from intra-class-switch move, distribution family has also been changed in inter-class-switch move ($k'\neq k$) as well as scale and shape parameters. Candidate distribution families are equiprobable for the candidate set $\{ 1, 2, 3\}\backslash\{k\}$, and we use functions below to propose candidate parameters of $\alpha'$ and $\gamma'$.

\begin{align}\label{equ:outswtransitions}
  \alpha' &= \psi(\alpha, k, k') \\
  \gamma' &= w(\alpha, \alpha', p, \gamma)
\end{align}

For intra-class transitions mentioned in the section above, the knowledge (about scale $\gamma$) learned in the previous algorithm steps was carried to the next step by the proposed method via FLOM based functions. The same approach is also utilized for $\gamma$ transitions in inter-class-switch move and functions $w(\cdot)$ are derived, however, this time, the sides of the transition are in different families. Details are shown in Table \ref{tab:LpInter}.

In order to perform efficient proposals for $\alpha$ in inter-class-switch move, instead of using a random move, we perform a mapping, $\psi(\cdot)$ from one family to another by taking into consideration the special members which are common for both of the families. For example, to derive an invertible mapping function on $\alpha$ for a transition from S$\alpha$S to Student's $t$, we utilize the information that Cauchy and Gauss distributions are common for both of the families. Cauchy refers to $\alpha=1$ for both of the families and Gauss refers to $\alpha=2$ for S$\alpha$S and $\alpha=\infty$ for Student's $t$. Hence, the invertible function $f_2(\alpha)$ performs the mapping for a transition from S$\alpha$S to Student's $t$.

Similarly, Gauss distribution is common for both S$\alpha$S and $\text{GG}$ for $\alpha$ value of 2. Thus, we derive another invertible function $f_1(\alpha)$ to move from S$\alpha$S to $\text{GG}$. Both of these mapping functions have been depicted in Figure \ref{fig:mappingfunc}.

$\text{GG}$ and Student's $t$ distributions have only Gauss distribution in common for $\alpha$ values of 2 and $\infty$, respectively. Due to having only one common distribution and infinite range of $\alpha$, instead of deriving an invertible mapping for transitions between these distributions, we perform a 2-stage mapping mechanism by firstly mapping $\alpha$ to S$\alpha$S from the most recent family, then mapping this value to the candidate family by using functions $f_1(\cdot)$ or $f_2(\cdot)$. Then the mapping from $\text{GG}$ to Student's $t$ is derived as: $\alpha' = f_2(f_1^{-1}(\alpha))$. It is straightforward to show that the reverse transition between shape parameters from Student's $t$ to GG results as $\alpha' = f_1(f_2^{-1}(\alpha))$. For all the transitions, mapping functions have been shown in Table \ref{tab:LpInter}.

So, the acceptance ratio for inter-class-switch move can be expressed as:

\begin{align}\label{equ:outswaccratio}
  A_{\text{inter-cl-sw}} = \min \left\{ 1, \dfrac{f(\mathbf{x}|k', \alpha', \gamma')}{f(\mathbf{x}|k, \alpha, \gamma)} \dfrac{f(\gamma')}{f(\gamma)} \dfrac{f(\alpha|k)}{f(\alpha'|k')} |J| \right\}
\end{align}
where $|J| = \dfrac{\partial \gamma'}{\partial \gamma} \dfrac{\partial \alpha'}{\partial \alpha}$.

\section{Experimental Study}\label{sec:sim}
We study experimentally three cases: synthetically generated noise, impulsive noise on PLC channels and 2-D DWT coefficients. Without loss of generality, distribution of data $\mathbf{x}$ is assumed to be symmetric around zero ($\delta=0$). The algorithm starts with a Gaussian distribution model with initial values $k^{(0)}=2$ and $\alpha^{(0)}=2$. Initial value for scale parameter $\gamma$ is selected as half of the interquartile range of the given data $\mathbf{x}$ and upper bounds $\alpha_{\text{max,S}\alpha\text{S}}, \alpha_{\text{max,GG}}$ and $\alpha_{\text{max},t}$ are selected as 2, 2 and 5, respectively. Some intuitive selections have been performed for the rest of the parameters. Move probabilities for intra-class-switch and inter-class-switch moves are assumed to be equally likely during the simulations. Additionally, in order to speed up the convergence of the distribution parameter estimations during the life move, which is the coefficient update move, it is chosen a bit more likely than intra-class-switch and inter-class-switch moves. Thus, the model move probabilities are selected as $P_{\text{life}}=0.4, P_{\text{intra-cl-sw}}=0.3$ and $P_{\text{inter-cl-sw}}=0.3$. Hyperparameters for prior distribution of $\gamma$ are set to $a=b=1$ and variance of proposal distribution for $\gamma$ in life move is set to $\xi_{scale} = 0.01$. Scale parameter $\Gamma$ of the discretized Laplace distribution for intra-class-switch move is selected as 0.4.

RJMCMC performs 5000 iterations in a single RJMCMC run and half of the iterations are discarded as burn-in period when estimating the distribution parameters. Random numbers from all the families have been generated by using Matlab's Statistics and Machine Learning Toolbox (for details please see\footnote{\url{https://www.mathworks.com/help/stats/continuous-distributions.html}}).

Performance comparison has been performed under two statistical significance tests, namely \textit{Kullback-Leibler} (KL) divergence and \textit{Kolmogorov-Smirnov} (KS) statistics. KL divergence has been utilized to measure fitting performance of the proposed method between estimated pdf and data histogram. Two-sample KS test compares empirical CDF of the data and the estimated CDF. It quantifies the distance between CDFs and performs an hypothesis test under a null hypothesis that two samples are drawn from the same distribution. Details about KL divergence and KS test have been discussed in Appendix \ref{appendixSSTests}.

\begin{table}[ht!]
  \centering
  \caption{Modeling results for synthetically generated processes.}\label{tab:syn}
  \small
  \resizebox{.7\linewidth}{!}{\begin{tabular}{ccccccc}
    \hline
    \textbf{Distribution} & \textbf{Est.} & \textbf{Est.} & \textbf{Est.} & \textbf{KL Div.} & \multicolumn{1}{l}{\textbf{KS}} & \multicolumn{1}{l}{\textbf{KS}}\\
    Distribution & \textbf{Family}& \textbf{Shape ($\hat{\alpha}$)} & \textbf{Scale ($\hat{\gamma}$)} & & \textbf{Score} & $p$\textbf{-value}\\
    \hline
    S$1.5$S$(2)$ & S$\alpha$S &1.4769 &	1.9162 &	0.0169 &	0.0125 &	1.0000\\
    S$1$S$(0.75)$ & $t$ &0.9970 &0.7300 &0.0454 &0.0489 & $>0.9999$\\
    GG$_{0.5}(0.5)$ & $\text{GG}$ &0.4990 &0.5199 &0.0229 &0.0152 & 1.0000\\
    GG$_{1.7}(1.4)$ & $\text{GG}$ &1.6456 &1.3374 &0.0221 &0.0202 & 1.0000\\
    $t_{3}(1)$ & $t$ &2.9303 &1.0039 &0.0251 &0.0203 & 1.0000\\
    $t_{0.6}(3)$ & $t$ &0.6197 &2.9869 &0.0465 &0.0452 & $>0.9999$\\
    \hline
    \end{tabular}}%
%\end{table}%
%\begin{table}[ht!]
  \centering
  \vspace{0.2cm}
  \caption{Modeling results for PLC impulsive noise.}\label{tab:PLC}
  \small
    \resizebox{.7\linewidth}{!}{\begin{tabular}{ccccccc}
    \hline
    \textbf{Data} & \textbf{Est.} & \textbf{Est.} & \textbf{Est.} & \textbf{KL Div.} & \multicolumn{1}{l}{\textbf{KS}} & \multicolumn{1}{l}{\textbf{KS}}\\
     & \textbf{Family}& \textbf{Shape ($\hat{\alpha}$)} & \textbf{Scale ($\hat{\gamma}$)} & & \textbf{Score} & $p$\textbf{-value}\\
    \hline
    \textit{PLC-1} &S$\alpha$S &1.2948 &5.6969 &0.0086 &0.0112 & 1.0000\\
    \textit{PLC-2} &S$\alpha$S& 0.7042 &0.1799 &0.0441 &0.0486 & $>0.9999$ \\
    \textit{PLC-3} &S$\alpha$S&1.3140 &1.3488 &0.0046 &0.0132 & 1.0000\\
    \hline
    \end{tabular}}%
%\end{table}%
%\begin{table}[ht!]
  \centering
  \vspace{0.2cm}
  \caption{Modeling results for 2D-DWT coefficients.}\label{tab:wave}
  \small
\resizebox{.7\linewidth}{!}{\begin{tabular}{lcccccc}
    \hline
    \textbf{Image} &\textbf{Est.} & \textbf{Est.} & \textbf{Est.} & \textbf{KL Div.} & \multicolumn{1}{l}{\textbf{KS}} & \multicolumn{1}{l}{\textbf{KS}}\\
     & \textbf{Family}& \textbf{Shape ($\hat{\alpha}$)} & \textbf{Scale ($\hat{\gamma}$)} & & Score & $p$-value\\
    \hline
    Lena (V)&$\text{GG}$    &0.5002 &1.7415 &0.0271 &0.0465 & $>0.9999$\\
    Lena (H)& $t$     &1.0958 &2.2422 &0.0094 &0.0349  & $>0.9999$\\
    Lena (D)& $t$     & 1.1628 &1.7735 &0.0145 & 0.0271  & 1.0000\\
          %&&&&&  \\
    SAR(V)     &S$\alpha$S   & 1.5381 &7.7395 &0.0025 &0.0123  & 1.0000\\
    SAR(H)     &S$\alpha$S   &1.4500 & 8.6249 &   0.0043 &        0.0221 & 1.0000\\
    SAR(D)    &       S$\alpha$S   &       1.7500 &       6.3710 &       0.0062 &       0.0125 & 1.0000 \\
          %&&&&&  \\
    MRI(V)     &       $\text{GG}$   &       0.3913 &      0.2693 &      0.0365 &      0.1152 & 0.8744 \\
    MRI(H)     &       $\text{GG}$    &       0.3527 &       0.1039 &       0.0305 &       0.0548  & $>0.9999$\\
    MRI(D)     &       S$\alpha$S   &       0.8504 &       0.5184 &       0.0245 &       0.0659 & $0.9998$ \\
          %&&&&&  \\
    Mammog.(V)     &       $t$     &       1.6325 &       1.6411 &       0.0363 &       0.0907 & 0.9816\\
    Mammog.(H)     &       $\text{GG}$    &       0.7501 &       1.5154 &       0.0121 &       0.0555 & $>0.9999$\\
    Mammog.(D)     &       $t$     &       1.6430 &       0.4851 &       0.0073 &       0.0117 & 1.0000 \\
          \hline
    \end{tabular}}%
\end{table}%

\begin{figure}[htbp]
\centering
\subfigure[S$1.5$S$(2)$]{%
\includegraphics[width=.8\linewidth]{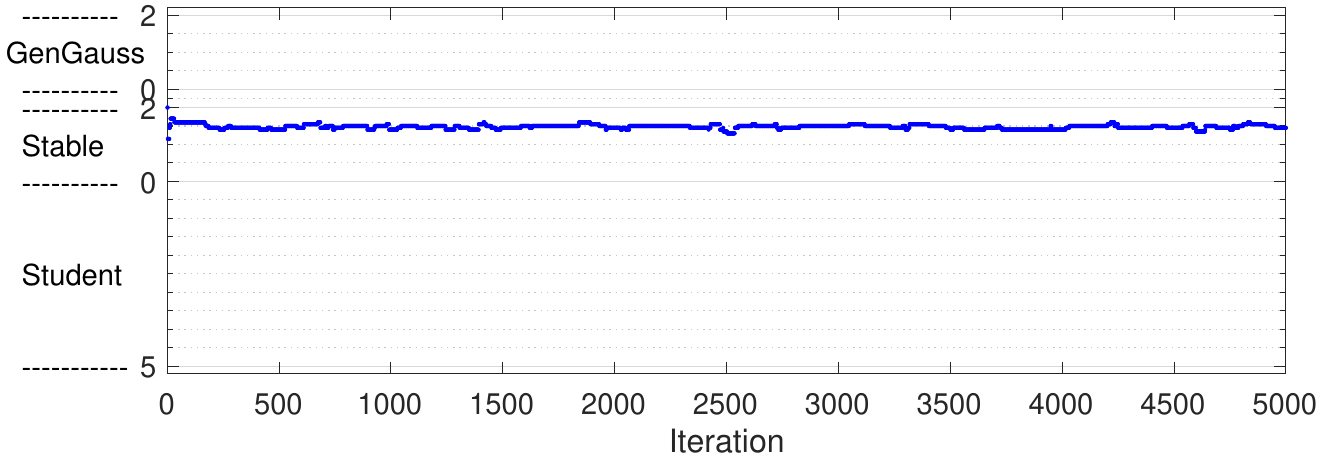}}
\centering
\hfil
\subfigure[GG$_{1.7}(1.4)$]{%
\includegraphics[width=.8\linewidth]{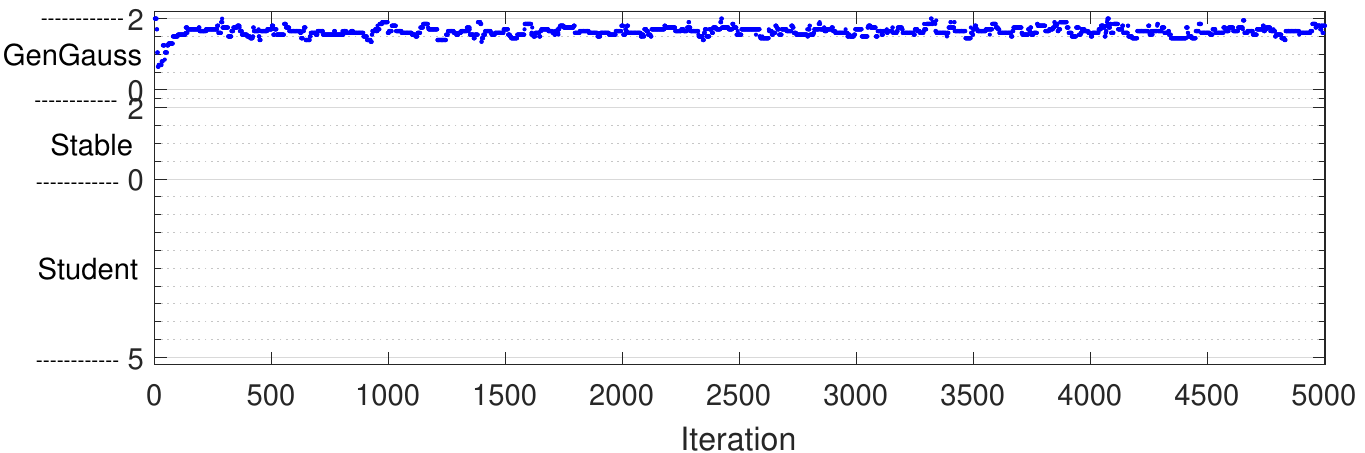}}
\hfil
\centering
\subfigure[$t_{3}(1)$]{%
\includegraphics[width=.8\linewidth]{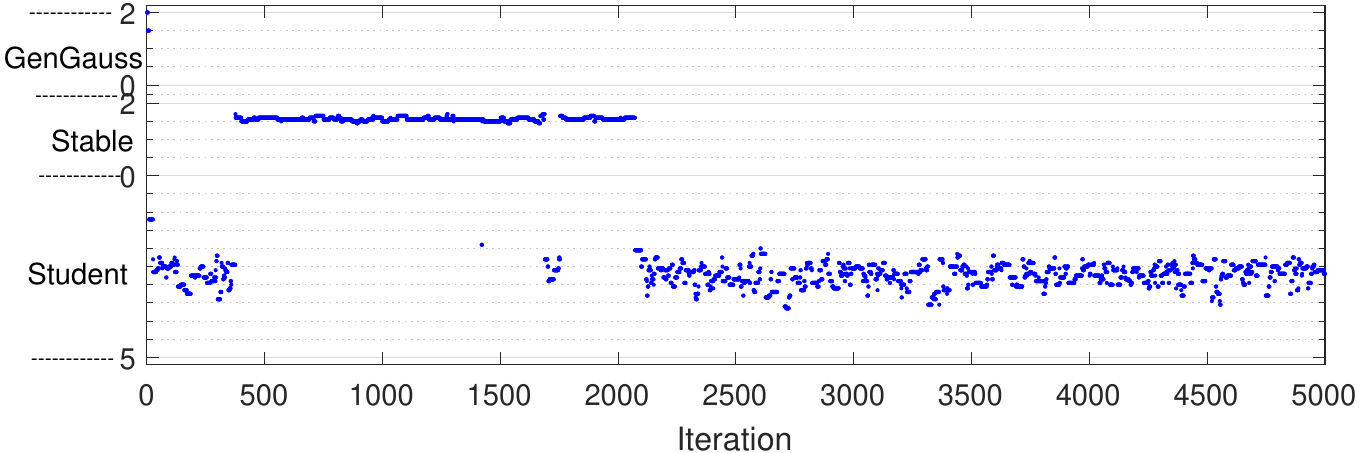}}
\centering
\hfil
\subfigure[S$1.5$S$(2)$]{
\includegraphics[width=.3\linewidth]{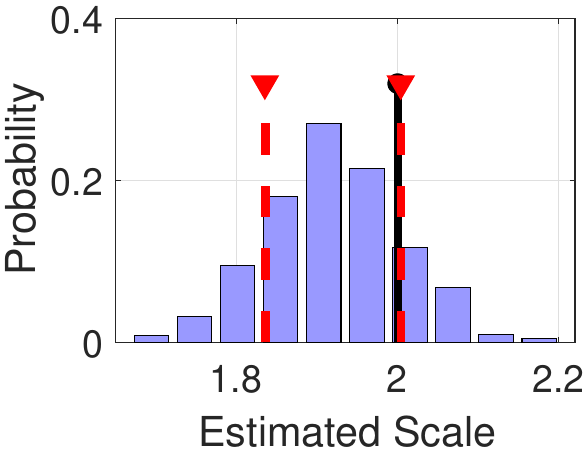}}
\centering
\hfil
\subfigure[GG$_{1.7}(1.4)$]{
\includegraphics[width=.3\linewidth]{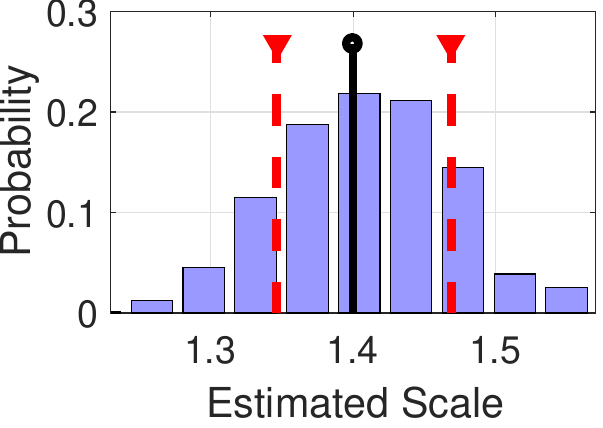}}
\centering
\hfil
\subfigure[$t_{3}(1)$]{
\includegraphics[width=.3\linewidth]{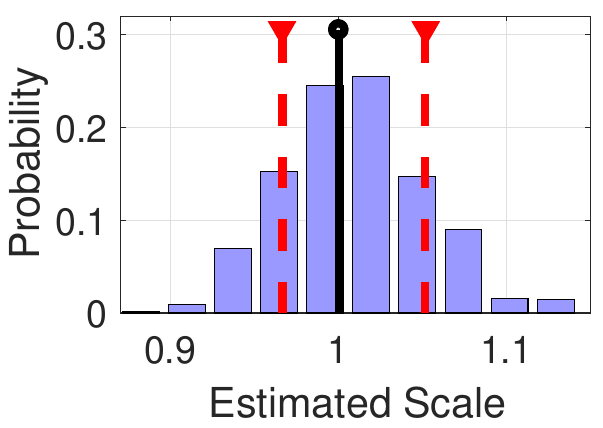}}
\hfil
\caption{Synthetically generated noise modeling - parameter estimation results in a single RJMCMC run. (a),(b),(c): Instantaneous $\alpha$ estimates. (d),(e),(f): Estimated posterior distributions for $\gamma$ after burn-in period.}
\label{fig:SynAll1}
\end{figure}

\subsection{Case Study 1: Synthetically Generated Noise Modeling}
In order to test the proposed method on modeling synthetically generated impulsive noise processes, six different distributions are chosen (2 distributions from each family). In a single RJMCMC run, data with a length of 1000 samples have been generated from one of the example distributions. The example distributions are S$1$S$(0.75)$, S$1.5$S$(2)$, GG$_{0.5}(0.5)$, GG$_{1.7}(1.4)$, $t_{3}(1)$ and $t_{0.6}(3)$.

40 RJMCMC runs have been performed for each distribution and estimated families with shape and scale parameters for each example distribution are shown in Table \ref{tab:syn}. In Figure \ref{fig:SynAll1}, instantaneous estimate of shape parameter $\alpha$ and estimated posterior distribution of scale parameter $\gamma$ are shown for three example distributions. Results represent the estimates obtained by a randomly selected RJMCMC run out of 40 runs. Burn-in period is not removed in the subfigures (a)-(c) in order to show the transient characteristics of the algorithm. These plots show that the proposed method converges to the correct shape parameters. In subfigures (d)-(f), vertical dashed-lines with $\nabla$ markers refer to $\pm\sigma$ \emph{confidence interval} (CI). Examining these subfigures shows that correct scale parameters lie within the $\pm\sigma$ CI of the posteriors.

\begin{figure*}[htbp]
\centering
\subfigure[S$1.5$S$(2)$]{%
\includegraphics[width=.4\linewidth]{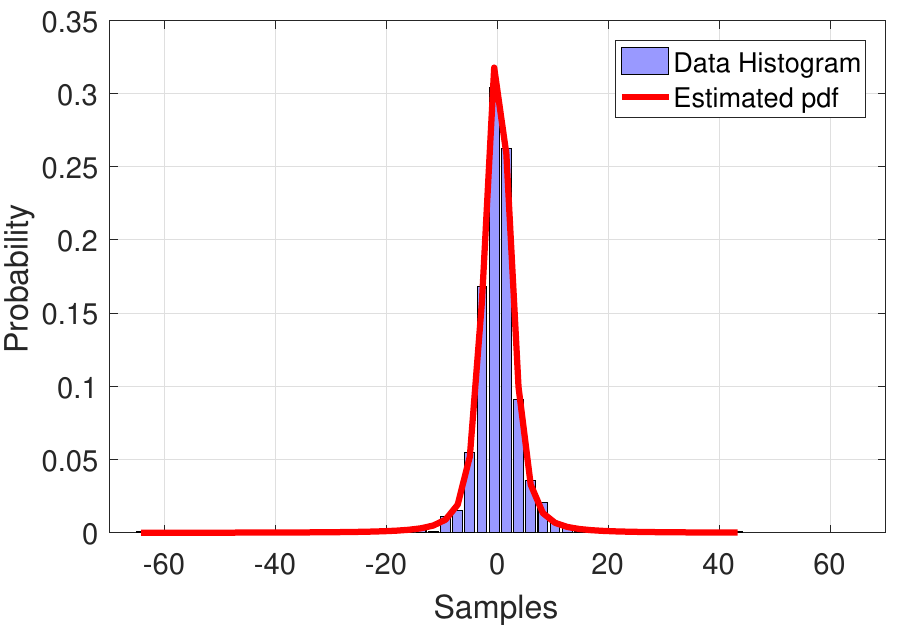}}
\centering
\subfigure[S$1.5$S$(2)$]{%
\includegraphics[width=.4\linewidth]{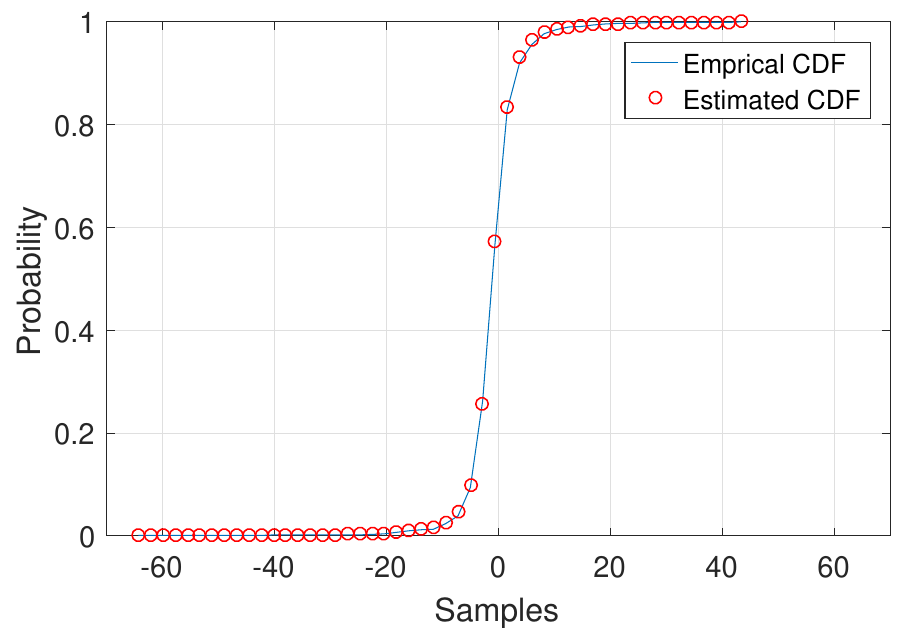}}
\centering
\subfigure[GG$_{1.7}(1.4)$]{
\includegraphics[width=.4\linewidth]{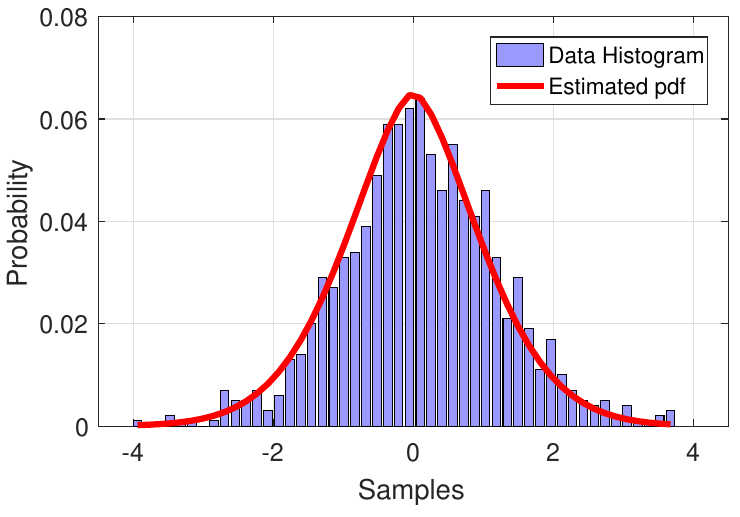}}
\centering
\subfigure[GG$_{1.7}(1.4)$]{
\includegraphics[width=.4\linewidth]{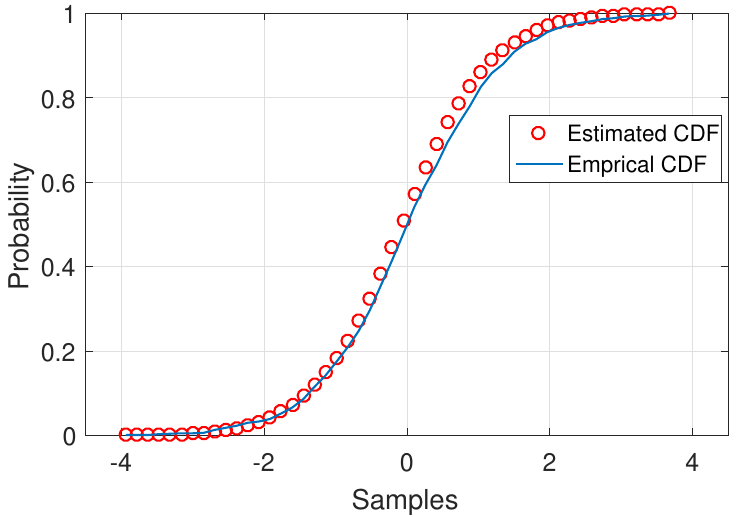}}
\centering
\subfigure[$t_{3}(1)$]{%
\includegraphics[width=.4\linewidth]{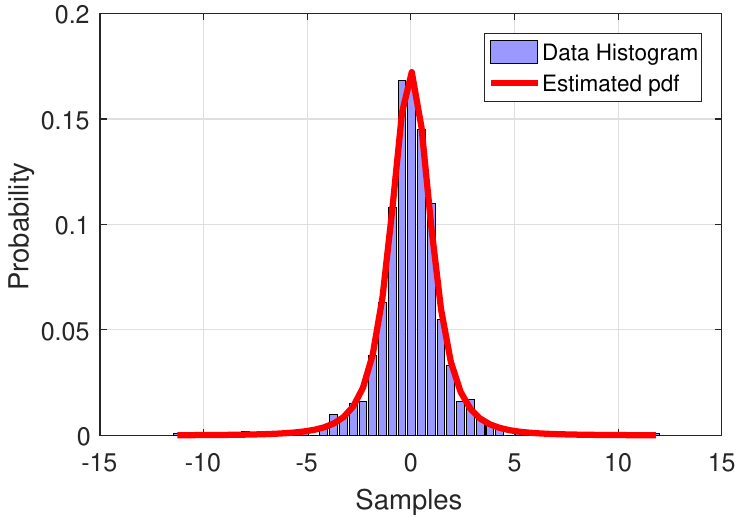}}
\centering
\subfigure[$t_{3}(1)$]{%
\includegraphics[width=.4\linewidth]{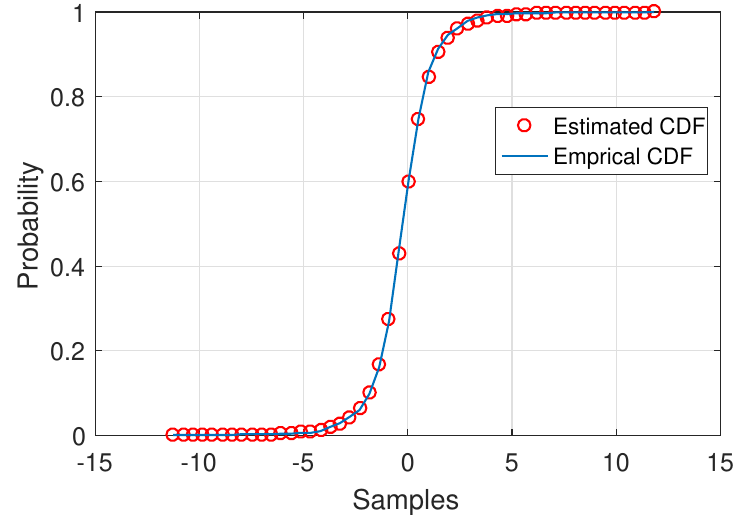}}
\caption{Synthetically generated noise modeling results. (a)-(c): Estimated pdfs, (d)-(f): Estimated CDFs.}
\label{fig:SynAll2}
\end{figure*}

Estimated pdfs and CDFs for three example distributions are depicted in Figure \ref{fig:SynAll2}. In addition to the statistical significance values in Table \ref{tab:syn}, fitting performance of the algorithm has been presented visually. As can be seen in Figure \ref{fig:SynAll2}, estimated pdfs are very similar to the data histogram and fitting performances for all example distributions lie within KL distance of at most 0.0465. Moreover, estimated CDFs under KS statistic score are also very low and $p$-values are close to 1,0000. Please note that the estimation result in the second line of Table \ref{tab:syn} is meaningful for an example Cauchy distribution, since the Cauchy distribution is a special member in both S$\alpha$S and Student's $t$ families.

%\section{Real Applications}\label{sec:RealApps}
\subsection{Case Study 2: Modelling Impulsive Noise on PLC Systems}
\begin{figure*}[ht!]
\centering
\subfigure[PLC-1]{\label{fig:PLCtime1}
\includegraphics[width=.31\linewidth]{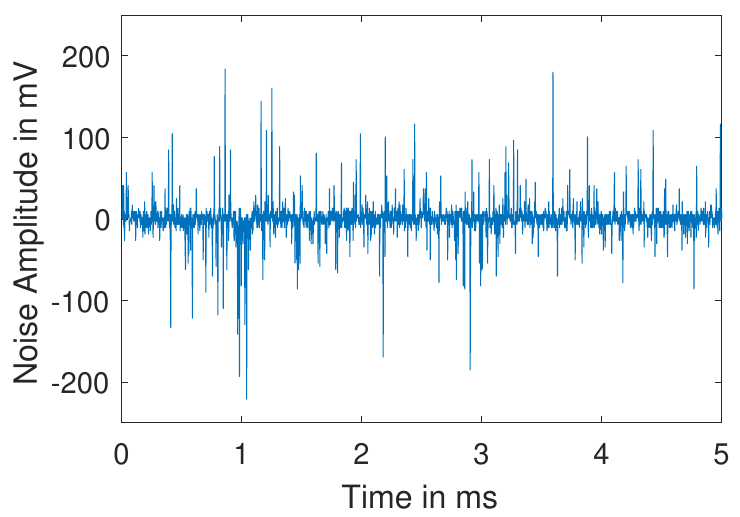}}
\subfigure[PLC-2]{\label{fig:PLCtime2}
\includegraphics[width=.31\linewidth]{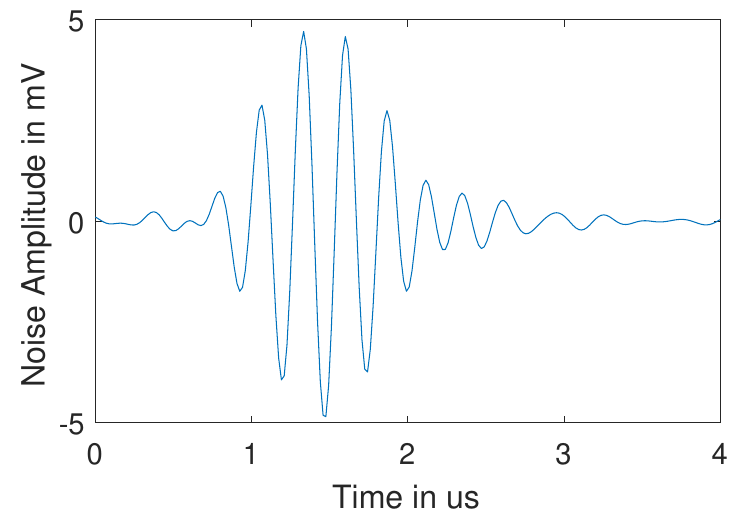}}
\subfigure[PLC-3]{\label{fig:PLCtime3}
\includegraphics[width=.31\linewidth]{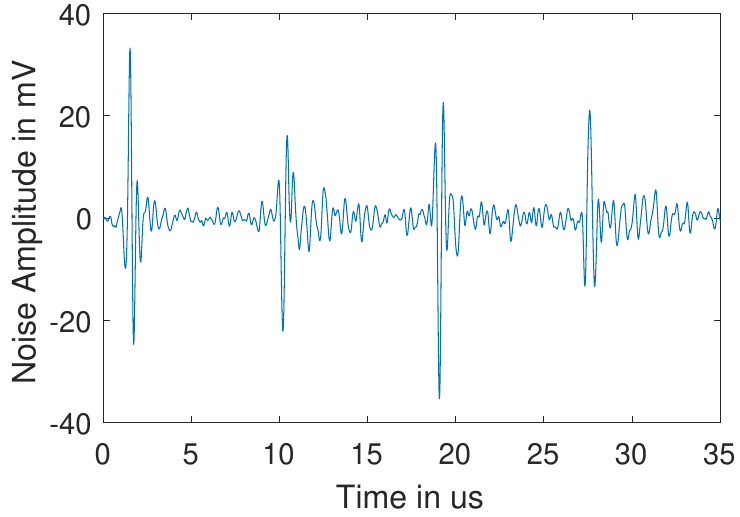}}
\hfil\\
\centering
\subfigure[PLC-1]{\label{fig:PLCfit1}
\includegraphics[width=.31\linewidth]{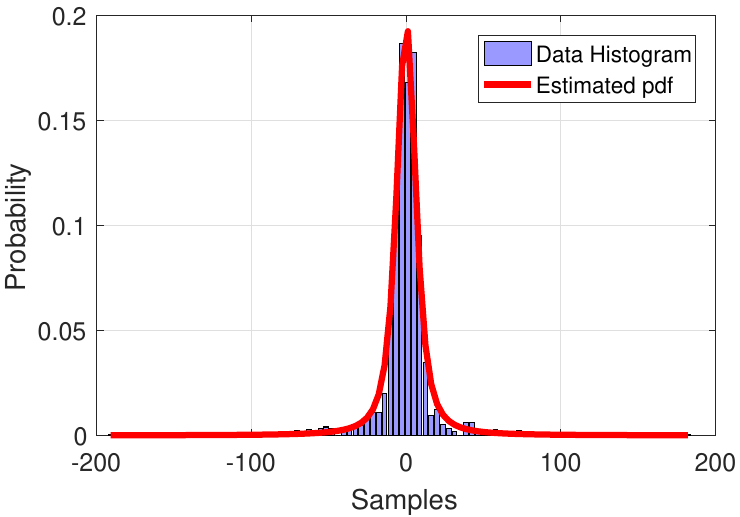}}
\subfigure[PLC-2]{
\includegraphics[width=.31\linewidth]{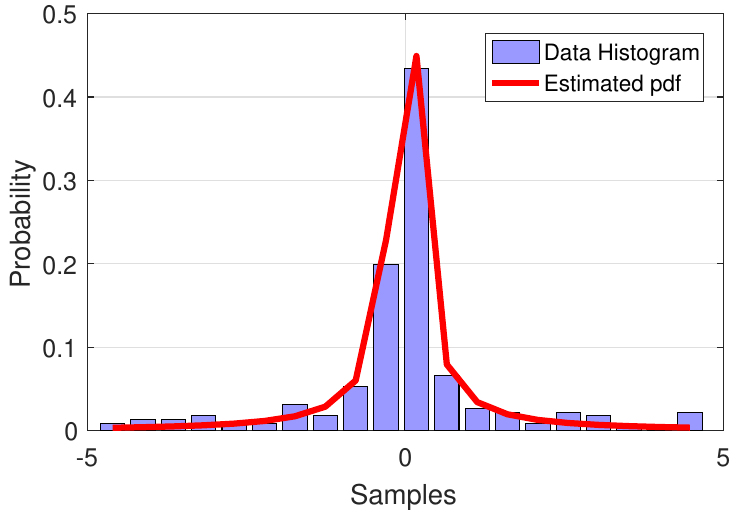}}
\subfigure[PLC-3]{
\includegraphics[width=.31\linewidth]{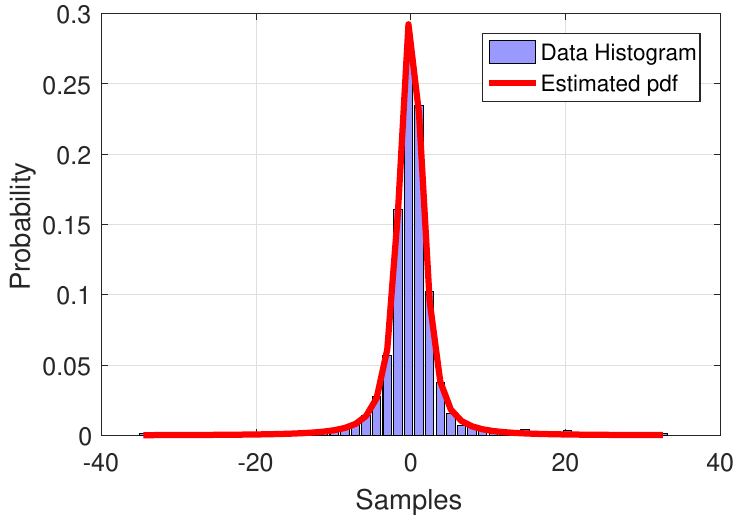}}
\hfil\\
\subfigure[PLC-1]{%
\includegraphics[width=.31\linewidth]{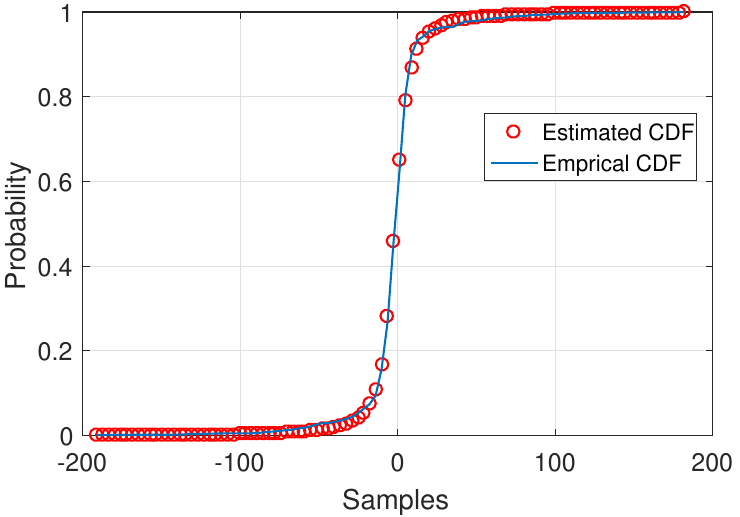}}
\subfigure[PLC-2]{
\includegraphics[width=.31\linewidth]{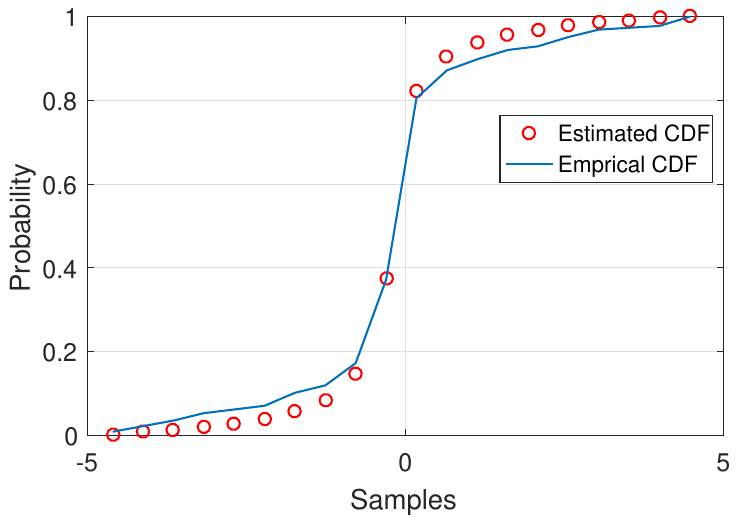}}
\subfigure[PLC-3]{\label{fig:PLCfit2}
\includegraphics[width=.31\linewidth]{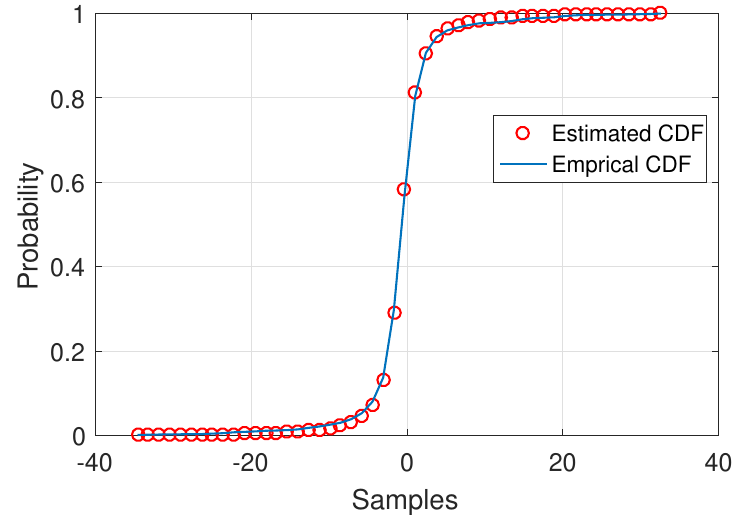}}
\hfil
\caption{PLC impulsive noise modeling results. (a)-(c): Time plots, (d)-(f): Estimated pdfs, (g)-(i): Estimated CDFs.}
\label{fig:PLCAll}
\end{figure*}
PLC is an emerging technology which utilizes power-lines to carry telecommunication data. Telecommunication speeds up to 200 Mb/s with a good quality of service can be achieved on PLC systems. Apart from this, PLC offers a physical medium for indoor multimedia data traffic without additional cables \cite{laguna2015use}.

A PLC system has various types of noise arising from electrical devices connected to power line and external effects via electromagnetic radiation, etc. These noise sequences are generally non-Gaussian and they are classified into three groups, namely: i) Impulsive noise, ii) Narrowband noise, iii) Background Noise \cite{cortes2010analysis}. Among these, impulsive noise is the most common cause of decoding (or communications) error in PLC systems due to its high amplitudes up to 40 dBs \cite{andreadou2010modeling}.

In this case study, we are going to use 3 different PLC noise measurements. First measurement (named as \emph{PLC-1}) has been performed during a project with number PTDC/EEA-TEL/67979/2006. Details for the measurement scheme and other measurements please see \cite{lopes2013dealing}. Data utilized in this thesis (\emph{PLC-1}) is an amplified impulsive noise measurement from a PLC system with a sampling rate of 200Msamples/sec. Measurements last for 5ms and there are 100K samples in the data set. In order to reduce the computational load, the data is downsampled with a factor of 50 and the resulting 2001 samples have been used in this study. In Figure \ref{fig:PLCAll}-(a) a time plot of the utilized downsampled data is depicted (For detailed description of the data please see\footnote[2]{http://sips.inesc-id.pt/$\sim$pacl/PLCNoise/index.html}).

Remaining two data sets are periodic synchronous and asynchronous (named as \emph{PLC-2} and \emph{PLC-3}, respectively) impulsive noise measurements both of which have been performed during project with number TIC2003-06842 (for details please see \cite{cortes2010analysis}). Periodic synchronous measurements last for 4$\mu$s and contain 226 noise samples. Periodic asynchronous measurements contain 1901 noise samples and last for 35$\mu$s. In Figures \ref{fig:PLCAll} (b) and (c) time plots are depicted for synchronous and asynchronous noise sequences, respectively (For detailed description of the data please see\footnote[3]{http://www.plc.uma.es/channels.htm}).

RJMCMC has been run 40 times for all three data sets. In Table \ref{tab:PLC}, estimated distribution families and resulting scale and shape parameters are depicted with significance test results. Estimated scale and shape parameters correspond to the average values after 40 repetitions. Examining the results in Table \ref{tab:PLC}, we can state that all three considered PLC noise processes follow S$\alpha$S distribution characteristics. In the literature, there are studies \cite{laguna2015use, tran2013plc} which model the impulsive noise in PLC systems by using stable distributions. Particularly, these studies provide a direct modelling scheme via stable distribution, whereas the proposed method has estimated the distribution among three impulsive distribution families. Thus, our estimation results for impulsive noise in PLC systems provide experimental verification to these studies. According to the results of KL and KS statistics shown in Table \ref{tab:PLC} on estimated pdfs and CDFs and Figures between \ref{fig:PLCfit1} and \ref{fig:PLCfit2}, RJMCMC fits to real data with a remarkable performance. KS $p$-values are all approximately 1 ($>0.9999$) and this provides strong evidence that the estimated and the correct distributions are of the same kind.

\begin{figure*}[ht!]
\centering
\subfigure[Lena]{
\includegraphics[width=.25\linewidth]{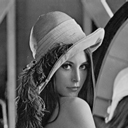}}
\hfil
\centering
\subfigure[Lena - Coefficient H]{
\includegraphics[width=.35\linewidth]{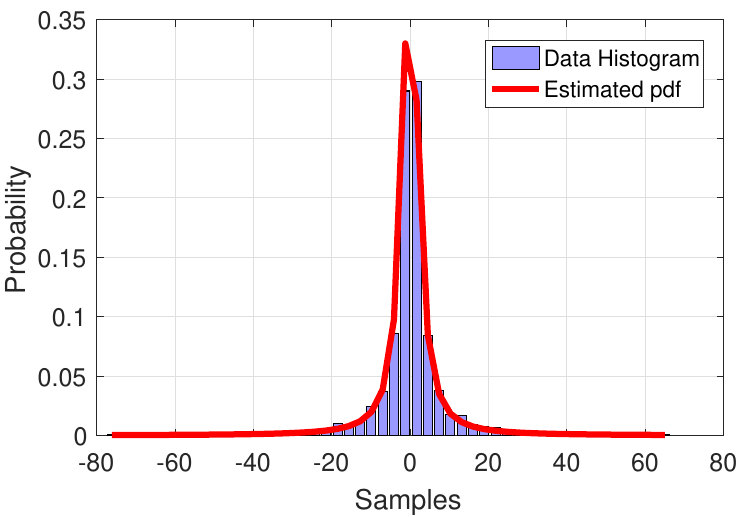}}
\hfil
\centering
\subfigure[Lena - Coefficient H]{%
\includegraphics[width=.35\linewidth]{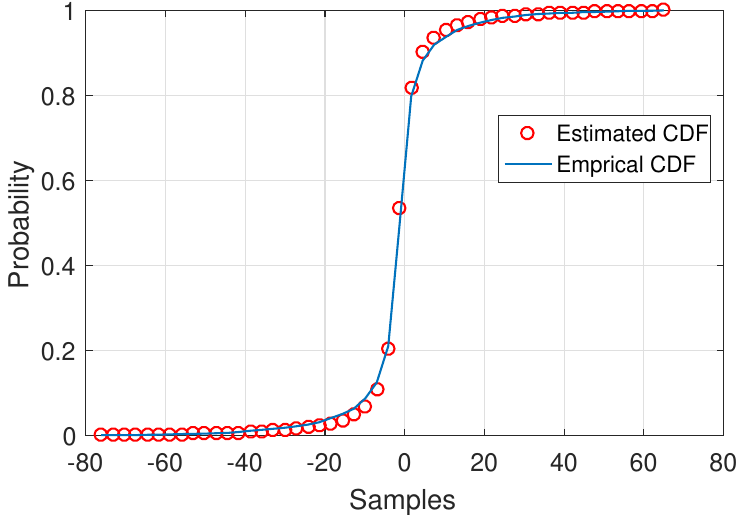}}
\hfil\\
\centering
\subfigure[SAR \cite{SAR3}]{
\includegraphics[width=.25\linewidth]{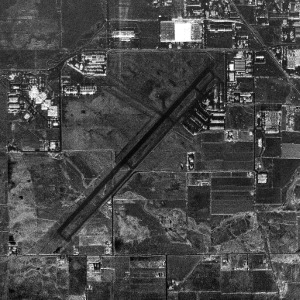}}
\hfil
\centering
\subfigure[SAR - Coefficient D]{
\includegraphics[width=.35\linewidth]{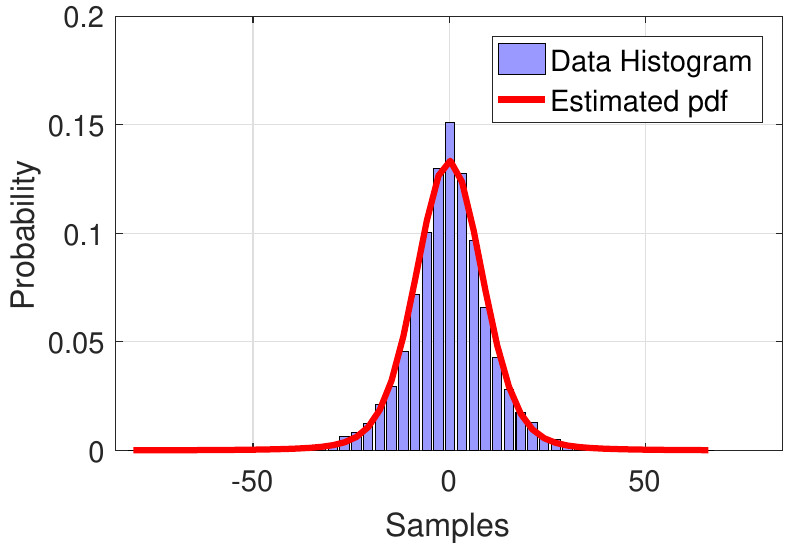}}
\hfil
\centering
\subfigure[SAR - Coefficient D]{
\includegraphics[width=.35\linewidth]{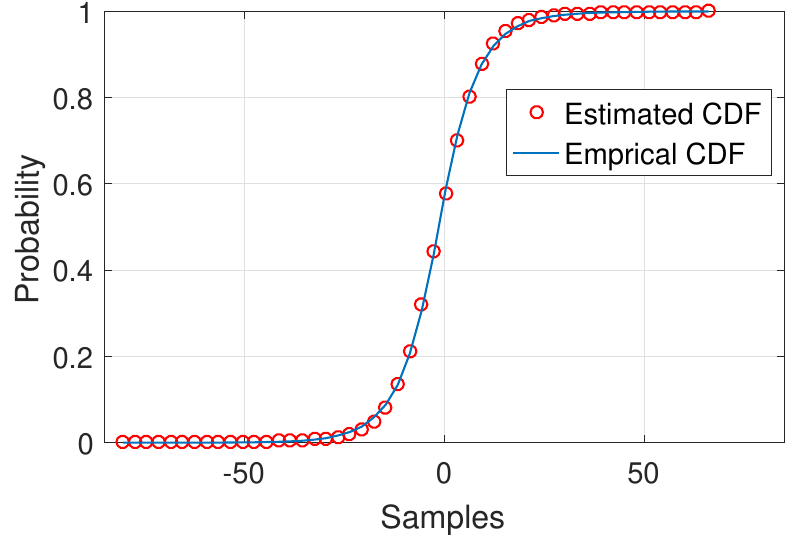}}
\hfil
\caption{2D-DWT coefficients modeling results for Lena and SAR images. (a),(d): Images, (b)-(e): Estimated pdfs, (c)-(f): Estimated CDFs.}
\label{fig:Images21}
\end{figure*}

\subsection{Case Study 3: Statistical Modeling for Discrete Wavelet Transform (DWT) Coefficients}
DWT which provides a multiscale representation of an image is a very important tool for recovering local and non-stationary features in an image. The resulting representation is closely related with the processing of the human visual system. DWT obtains this multiscale representation by performing a decomposition of the image into a low resolution approximation and three detail images capturing horizontal, vertical and diagonal details. It has been observed by several researchers that they have more heavier tails and sharper peaks than Gaussian distribution \cite{simoncelli1997statistical, achim2003sar}.%, achim2001novel, portilla2003image}.

\begin{figure*}[ht!]
\centering
\subfigure[MRI \cite{MRI}]{%
\includegraphics[width=.25\linewidth]{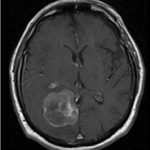}}
\centering
\subfigure[MRI - Coefficient V]{%
\includegraphics[width=.35\linewidth]{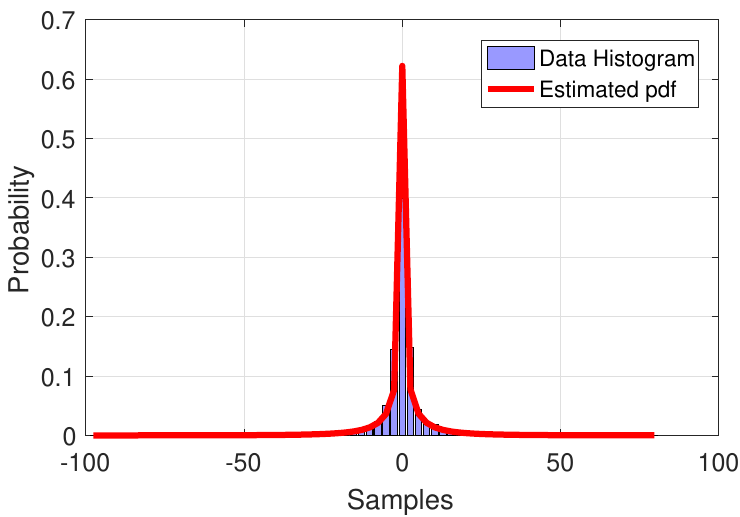}}
\hfil
\centering
\subfigure[MRI - Coefficient V]{%
\includegraphics[width=.35\linewidth]{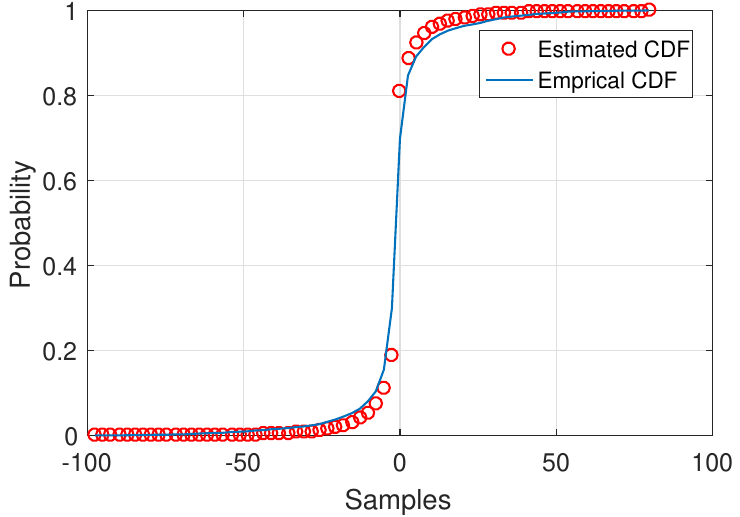}}
\hfil\\
\centering
\subfigure[Mammogram \cite{mammogram1}]{%
\includegraphics[width=.25\linewidth]{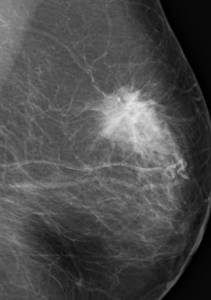}}
\hfil
\centering
\subfigure[Mammogram - Coefficient D]{%
\includegraphics[width=.35\linewidth]{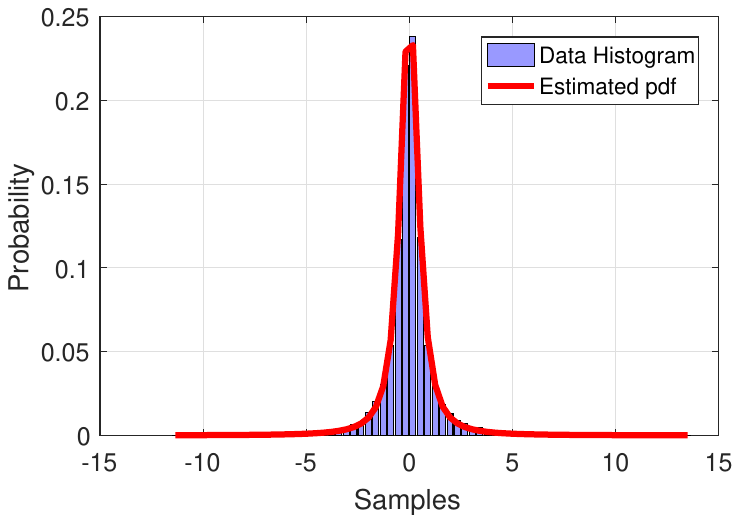}}
\hfil
\centering
\subfigure[Mammogram - Coefficient D]{%
\includegraphics[width=.35\linewidth]{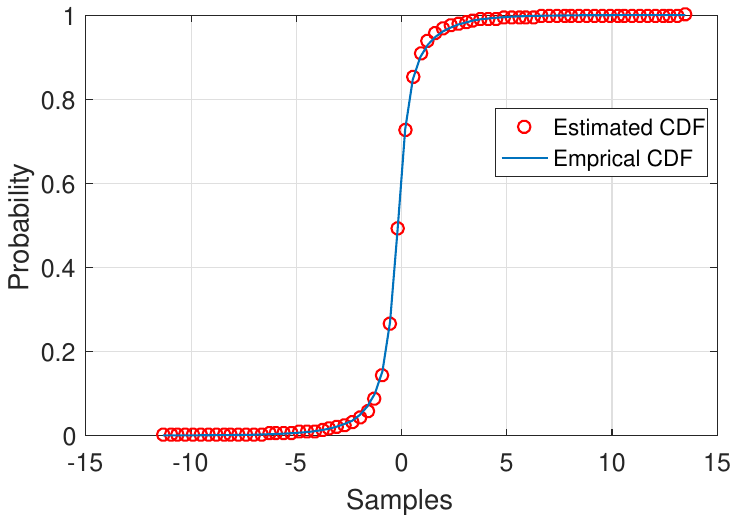}}
\hfil
\caption{2D-DWT coefficients modeling results for MRI and Mammogram. (a),(d): Images, (b)-(e): Estimated pdfs, (c)-(f): Estimated CDFs.}
\label{fig:Images22}
\end{figure*}

In this study, the proposed method has been utilized to model the coefficients (e.g. subbands) of 2D-DWT, namely vertical (V), horizontal (H) and diagonal (D). Four different images have been used to test the performance of the algorithm under statistical significance tests: Lena, \textit{synthetic aperture radar} (SAR) \cite{SAR3}, \textit{magnetic resonance imaging} (MRI) \cite{MRI} and mammogram \cite{mammogram1} which are shown in the first columns of Figures \ref{fig:Images21} and \ref{fig:Images22}.

The proposed method has been performed for 40 RJMCMC runs. Estimated results for distribution families and their parameters ($\alpha$ and $\gamma$) are depicted in Table \ref{tab:wave} as averages of 40 runs.

Estimated distributions for wavelet coefficients of images in Table \ref{tab:wave} show different characteristics. SAR and MRI images follow generally S$\alpha$S characteristics while results for Lena and mammogram images are generally $\text{GG}$ or Student's $t$. Moreover, despite modelling with different distribution families, all the coefficients for all the images have been modelled successfully according to the KL and KS test scores and $p$-values. The estimated pdfs and CDFs in Figures \ref{fig:Images21} and \ref{fig:Images22} show remarkably good fitting and provide support to the results which are obtained numerically in Table \ref{tab:wave}.

\subsection{Graphical Evaluation by Q-Q Plots for Data Estimated as S$\alpha$S}
Quantile-Quantile plot, or simply Q-Q plot can be described as a graphical representation of the sorted quantiles of a data set against the sorted quantiles of another data set. Suppose we have two samples with length $n\text{,} X_1, X_2, \ldots, X_n$ and $Y_1, Y_2, \ldots, Y_n$. In terms of Q-Q plot, these two samples are from the same distribution, as long as their ordered sequences, $X_{(1)}, X_{(2)}, \ldots, X_{(n)}$ and $Y_{(1)}, Y_{(2)}, \ldots, Y_{(2)}$, should satisfy, $X_{(i)} \approx Y_{(i)} \quad i = 1, 2, \ldots, n.$.

Q-Q has been used to compare distributions of two populations, or compare distribution of one population to a reference distribution. Q-Q plot may provide information about the location, shape and scale parameters comparison between two populations.

\begin{figure}[ht!]
\centering
\subfigure[S1.5S(2)]{%
\includegraphics[width=0.32\linewidth]{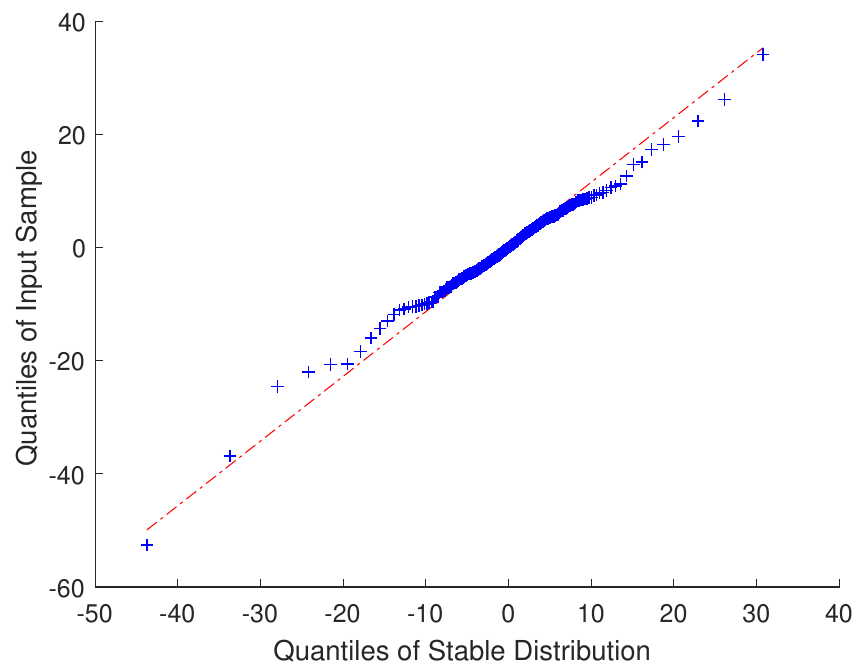}}
\centering
\subfigure[\textit{PLC-1}]{
\includegraphics[width=0.32\linewidth]{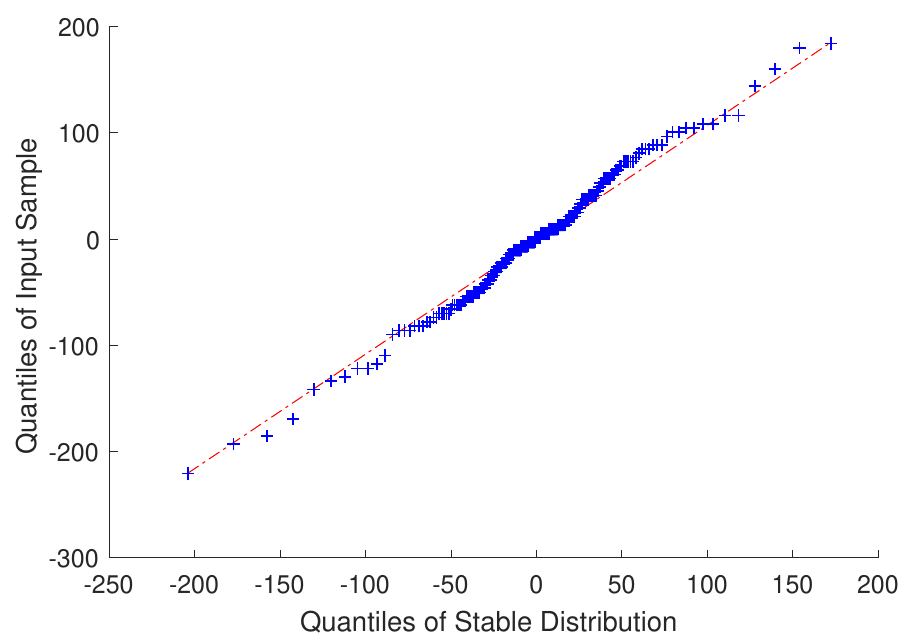}}
\centering
\subfigure[\textit{PLC-2}]{%
\includegraphics[width=0.32\linewidth]{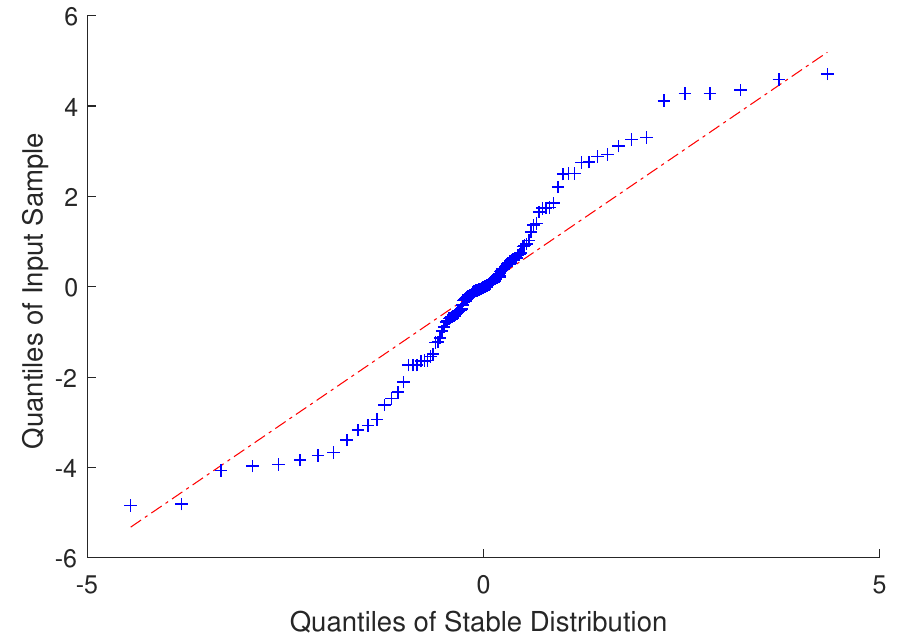}}
\centering
\subfigure[\textit{PLC-3}]{
\includegraphics[width=0.32\linewidth]{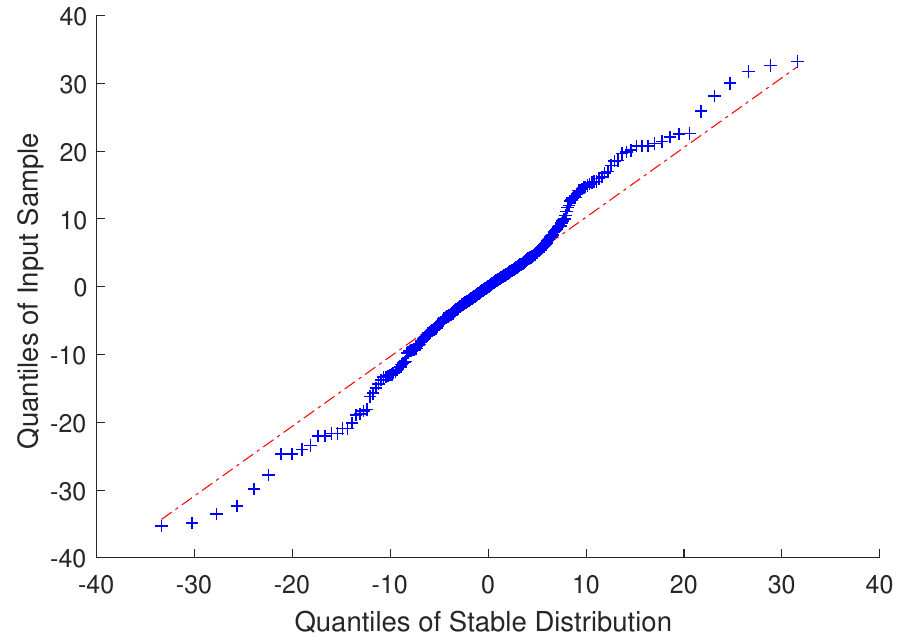}}
\centering
\subfigure[SAR-V]{%
\includegraphics[width=0.32\linewidth]{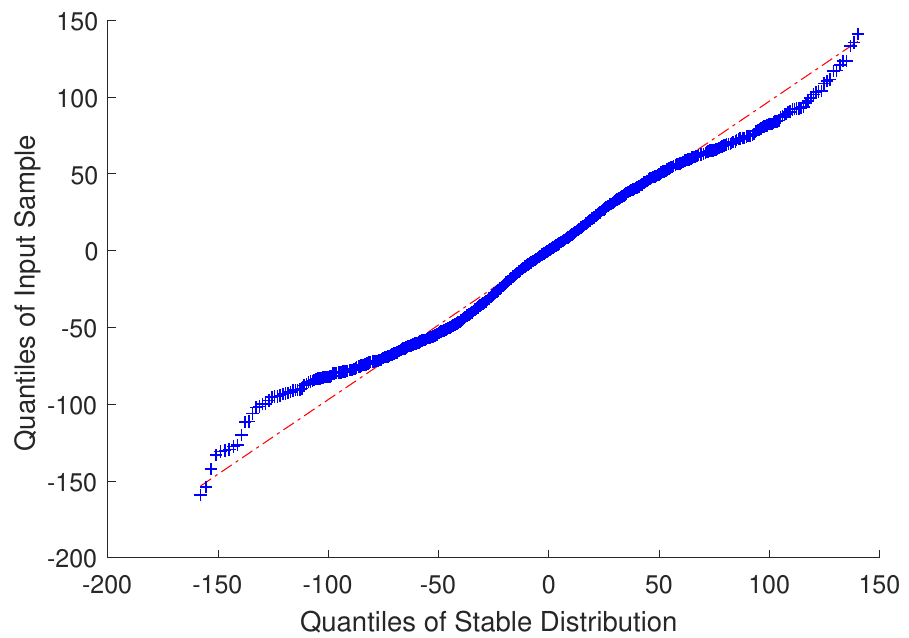}}
\centering
\subfigure[SAR-H]{
\includegraphics[width=0.32\linewidth]{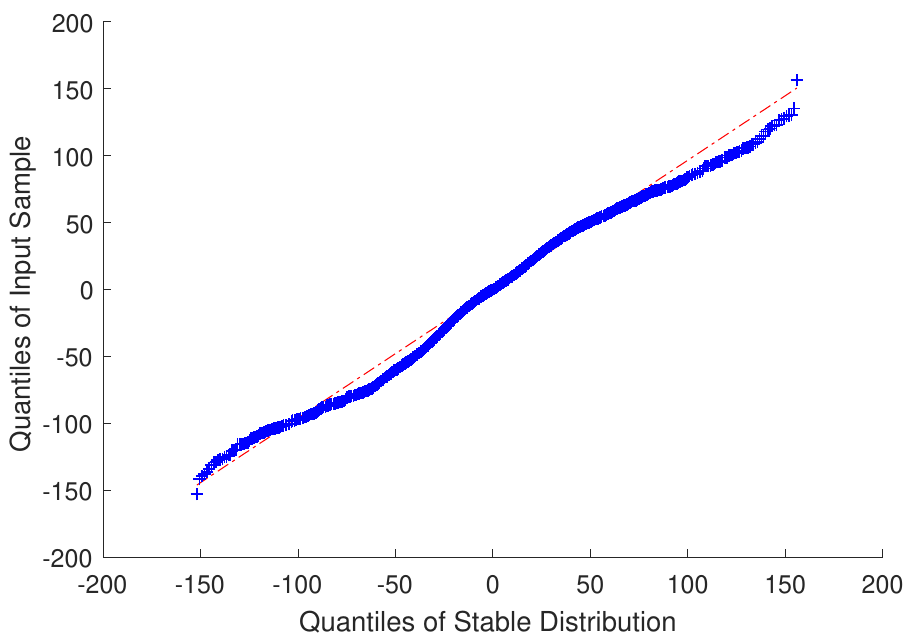}}
\centering
\subfigure[SAR-D]{%
\includegraphics[width=0.32\linewidth]{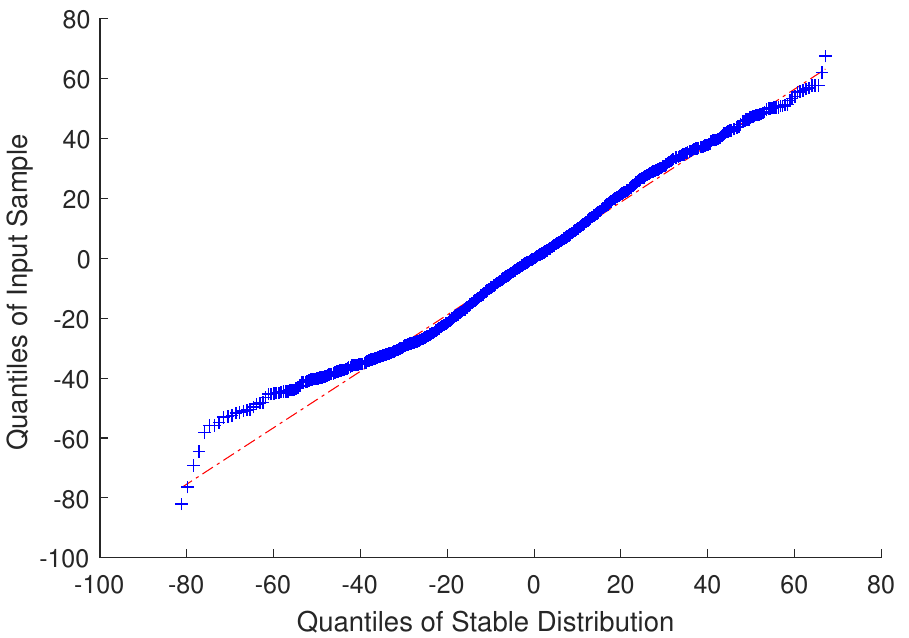}}
\centering
\subfigure[MRI-D]{
\includegraphics[width=0.32\linewidth]{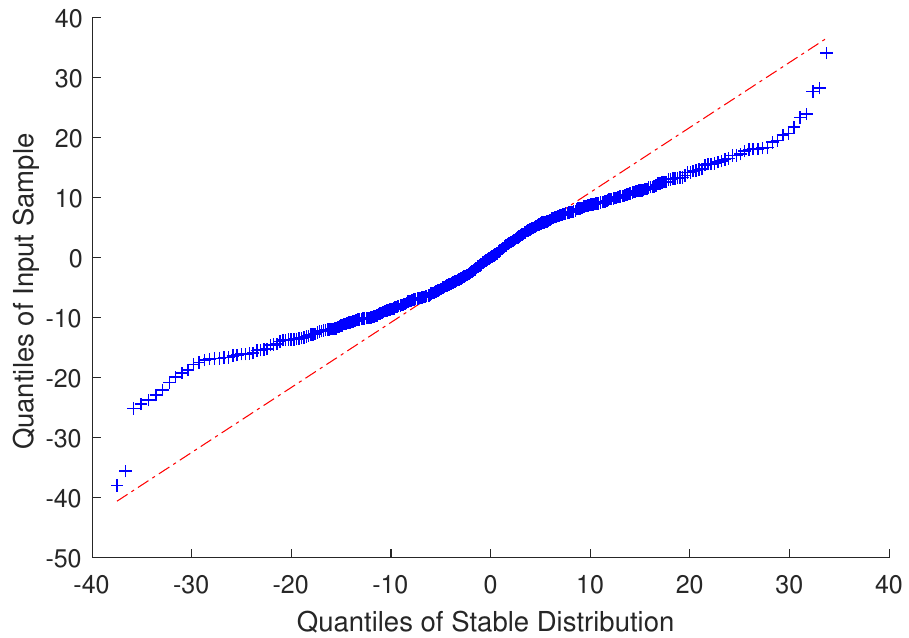}}
\caption{Q-Q plots for S$\alpha$S estimated data sets.}
\label{fig:Appqq1}
\end{figure}

Figure \ref{fig:Appqq1} shows Q-Q plots for data sets estimated to be S$\alpha$S. Examining the figures clearly shows a remarkable match between estimated distribution and the data samples. Q-Q plots for PLC-2 and MRI-D results in Figure \ref{fig:Appqq1}-(c) and (h), respectively, show relatively lower performance than the others. However, this result is expected because the numerical estimation results in terms of KL and KS scores obtained for these two data sets are already higher than the others (KS scores of 0.0486 and 0.0659, respectively), but still acceptable due to $p$-values of 0.9999 and 0.9998, respectively.

\section{Conclusion}\label{sec:conclusion}
In this study, we have utilized RJMCMC beyond the framework of trans-dimensional sampling which we call trans-space RJMCMC. By defining a new combined parameter space of current and target parameter subspaces of possibly different classes or structures, we have shown that the original formulation of RJMCMC offers more general applications than just estimating the model order. This provides users to do model selection between different classes or structures. In particular, exploring solution spaces of linear and nonlinear models or of various distribution families is possible using RJMCMC. One can expect higher benefits from the trans-space RJMCMC compared to considering different model classes separately in the cases when the different model class spaces have intersections to exploit. The intersections for the trans-distributional RJMCMC considered in this paper have been the common distributions in the impulsive noise families. They made it possible to use the mapping functions benefiting from the FLOMs of the observed data. These functions in turn have enabled to transfer the information learned while searching in one family to the subsequent search after an inter-class-switch move.

Candidate distribution space covers various impulsive densities from three popular families, namely S$\alpha$S, $\text{GG}$ and Student's $t$. In both synthetically generated noise processes and real PLC noise measurements and wavelet transforms of images, the proposed method shows remarkable performance in modeling. Simulation studies verify the remarkable performance in modelling the distributions in terms of both visual and numerical tests. KL and KS tests show the numerical results are statistically significant in terms of $p$-values which are generally close to 1.0000 (at least 0.85) for all the example data sets. Moreover, the algorithm indicated S$\alpha$S distributions for 2D-DWT coefficients of SAR images and noise on PLC channels which is in accordance with the other studies in the literature and confirms the success of the algorithm.

We would like to underline that the ideas presented in this paper are not limited only to sampling across distribution families but can be extended to any class of models.

\section*{Acknowledgement}
Oktay Karakuş is funded as a visiting scholar at ISTI-CNR, Pisa, Italy by The Scientific and Technological Research Council of Turkey (TUBITAK) under grant program 2214/A.

\appendix
\section{Statistical Significance Tests}\label{appendixSSTests}
\subsection{Kullback-Leibler Divergence}
In order to measure the difference between two probability distributions, in probability and statistics, a well-known approach named \textit{Kullback-Leibler divergence} or shortly \textit{KL divergence} has been commonly used. KL divergence provides a non-symmetric measure about how different two probability distributions, e.g. $p$ and $g$ are, and also known as relative entropy between $p$ and $g$.

KL divergence between two continuous probability distributions $p(x)$ and $g(x)$ can be defined as:\begin{eqnarray}
D_{KL}(p\|q) = E\left[ \log(p(x)) - \log(g(x)) \right],
\end{eqnarray}
where $\log(\cdot)$ refers to the natural logarithm, and $E[\cdot]$ is the expectation. The most common way to represent $D_{KL}(p\|q)$ is \cite{kullback1997information,hershey2007approximating}:\begin{eqnarray}\label{equ:KLcont}
D_{KL}(p\|q) = \int_x p(x) \log\left( \dfrac{p(x)}{g(x)} \right).
\end{eqnarray}
The notation $D_{KL}(p\|q)$ denotes ``\textbf{information lost where $g$ is used to approximate $p$}" \cite{burnham2003model}.

The KL divergence can also be used to measure the distance between discrete distributions, such as Poisson, negative binomial, or in cases when comparing two discrete populations. Discrete KL divergence can be defined as \cite{burnham2003model}:\begin{eqnarray}\label{equ:KLcont}
D_{KL}(p\|q) = \sum_{i=1}^{k} p(x_i) \log\left( \dfrac{p(x_i)}{g(x_i)} \right).
\end{eqnarray}
KL divergence satisfies three properties, which are \cite{hershey2007approximating}:
\begin{enumerate}
  \item Self-similarity $\rightarrow D_{KL}(p\|p) = 0$,
  \item Self-identification $\rightarrow D_{KL}(p\|q) = 0$ only if $p = g$,
  \item Positivity $\rightarrow D_{KL}(p\|q) \geq 0$ for all $p, g$.
\end{enumerate}

To understand the information that KL divergence provides clearly, let's create a toy example. Assume that a sequence of data, $x$ has been observed and the distribution of these samples, $f(x)$ will be tested to be uniform or binomial distributions which refer to $f_1$ and $f_2$, respectively. The equation has been used and KL divergence values are calculated. Resulting values are, $D_{KL}(f\|f_1) = 0.22$ and $D_{KL}(f\|f2) = 0.105$. Examining the KL divergence values, we can state that the distribution of observed samples is more likely to come from uniform distribution, or conversely approximating the distribution of the observed samples with binomial distribution causes more information loss than uniform distribution.

\subsection{Kolmogorov-Smirnov Test}
Kolmogorov-Smirnov test, or simply KS test, can be defined as a non-parametric test which can be used to test equality of continuous distributions by comparing one sample with a reference distribution (one-sample KS test) or two samples (two-sample KS test).

Particularly, suppose that a population has a cumulative distribution function $F(x)$ (reference distribution) which is clearly specified. There is also an observed population the empirical cumulative distribution function of which is $G(x)$. One can think that a measure between these two distributions may provide means about whether the reference distribution is the correct one or not. KS test quantifies a measure for this purpose as \cite{massey1951kolmogorov,goodman1954kolmogorov,wilcox2005kolmogorov}:\begin{eqnarray}
D_{KS} = \max_x \left| F(x) - G(x) \right|,
\end{eqnarray}
If the calculated KS score is large, this provides evidence that the reference distribution $F(x)$ is not the correct distribution for observed samples. The measure $D_{KS}$ can be defined as one-sample KS score (or measure). If we deal with observations from two populations the empirical cumulative distribution functions of which are $F_1(x)$ and $F_2(x)$. Here, KS test will be used to test two populations come from the same distribution or not.  Two-sample KS test score is:\begin{eqnarray}
D_{KS} = \max_x \left| F_1(x) - F_2(x) \right|,
\end{eqnarray}

\begin{figure}[t]
    \centering
    \includegraphics[width=.7\linewidth]{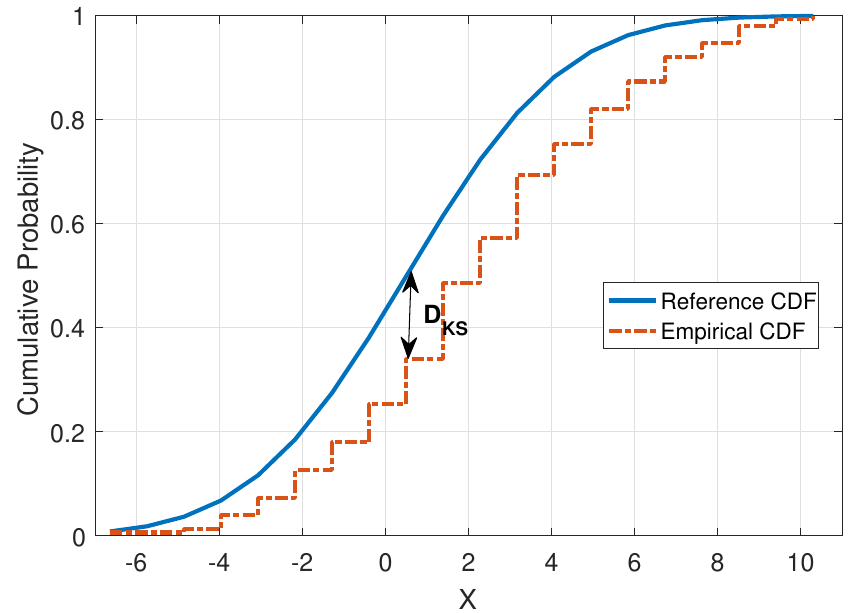}
    \caption{KS Score calculation example.}
    \label{fig:KS_ex}
\end{figure}

KS score has also meaning from a graphical point of view. The largest vertical distance between two cumulative distribution functions can be defined as KS test score \cite{wilcox2005kolmogorov}. In Figure \ref{fig:KS_ex}, an example to KS score is shown.

In addition to the KS test score explained above, KS test also performs a hypothesis testing under the null hypothesis that says ``two populations are drawn from the same underlying continuous population". This hypothesis, $\mathcal{H}$ is rejected providing that any given significance value, $\alpha$ is as large as or larger than $p$-value. In order to calculate $p$-value, the limiting forms of the Kolmogorov's distribution should be calculated \cite{massey1951kolmogorov,wang2003evaluating,press2007numerical}:\begin{eqnarray}
\lim_{n\to\infty} Pr(D_{KS} \leq t) = L(t) = 1 - 2\sum_{i=1}^{\infty} (-1)^{i-1} \exp(-2i^2t^2).
\end{eqnarray}
The corresponding $p$-value can be computed as \cite{press2007numerical,tong2010efficient}:\begin{eqnarray} \label{equ:pValAPP}
p\text{-value} = Pr(D_{KS} > t) = 1 - L(t) = 2\sum_{i=1}^{\infty} (-1)^{i-1} \exp(-2i^2t^2).
\end{eqnarray}
There is still one unknown, $t$, and it can be obtained approximately as \cite{stephens1970use,press2007numerical}:\begin{eqnarray} \label{equ:tAPP}
t = D_{KS}\left[N_e + 0.12 + \dfrac{0.11}{N_e}\right],
\end{eqnarray}
where $N_e$ refers to the sample size, $N$, for one-sample KS test and $\sqrt{\dfrac{N_1N_2}{N_1+N_2}}$ for two-sample KS test. Substituting $t$ obtained in (\ref{equ:tAPP}) with (\ref{equ:pValAPP}) gives $p$-value.

\bibliography{DensEst_biblio}

\end{document}